\documentclass[aps,prd,superscriptaddress,showpacs,twocolumn,preprintnumbers,showkeys,superscriptaddress,amsmath,amssymb,nofootinbib]{revtex4-2}
\usepackage{array}
\usepackage{booktabs}
\usepackage{commath}
\usepackage{graphicx}
\usepackage{epstopdf}
\usepackage{amsmath}
\usepackage{amsfonts}
\usepackage{amssymb}
\usepackage{latexsym}
\usepackage{hyperref}
\usepackage[english]{babel}
\usepackage[utf8]{inputenc}
\usepackage[colorinlistoftodos]{todonotes}
\usepackage{xcolor}
\usepackage{slashed}
\usepackage{bm}  
\usepackage[normalem]{ulem} 
\usepackage{amsmath,mathtools}
\usepackage{caption}
\usepackage{subcaption}
\usepackage{float}
\useunder{\uline}{\ul}{}
\usepackage{hyperref}
\usepackage{layout}
\begin{document}

\title{Reconciling LSND and super-Kamiokande data through the dynamical Lorentz symmetry breaking in a four-Majorana fermion model}

\author{Y. M. P. Gomes}
\email{yurimullergomes@gmail.com}
\affiliation{Departamento de F\'{\i}sica Te\'{o}rica, Universidade do
  Estado do Rio de
Janeiro, 20550-013 Rio de Janeiro, Brazil}

\author{M. J. Neves}\email{mariojr@ufrrj.br}
\affiliation{Departamento de F\'isica, Universidade Federal Rural do Rio de Janeiro,
BR 465-07, 23890-971, Serop\'edica, Rio de Janeiro, Brazil}




\begin{abstract}
We propose a model of Majorana fermions with quartic self-couplings. These Majorana fermions acquire masses via a type II seesaw mechanism in which the physical eigenstates are identified as a light Majorana fermion and another heavy Majorana fermion. On a physical basis, the quartic self-couplings involve axial currents of these Majorana fermions, and also the interaction of the axial current for the light particle
with the heavy particle one. We introduce two auxiliaries gauge fields in this model, and we study the stability conditions of the correspondent effective potential of the model. The ground state of the effective potential introduces two 4-vectors as scales of vacuum expected values, and consequently, the dynamical Lorentz symmetry breaking (DLSB) emerges in the model. We use the expansion of the effective action to calculate the effective Lagrangian up to second order in the auxiliary fields as fluctuations around the ground state. This mechanism generates dynamics for the auxiliary gauge fields, mixed mass terms, longitudinal propagation, and Chern-Simons term through radiative corrections. After the diagonalization, the two gauge fields gain masses through
an analogous type II seesaw mechanism in which a gauge boson has a light mass, and the other one acquires a heavy mass.
In this scenario of Lorentz symmetry breaking, we obtain the correspondent dispersion relations for the Majorana fermions and the
gauge boson fields. Posteriorly, we analyze the neutrino's oscillations in the presence of a DLSB parameter, in the transition $\nu_{e} \rightarrow \nu_{\mu}$. We discuss the parameter space of this transition and show that the DLSB can conciliate the LSND and super-Kamiokande results.

\end{abstract}

\pacs{14.60.Pq , 13.15.+g , 11.30.Cp, 11.15.Ex, 14.70.-e}

\keywords{Majorana fermions, Dynamical Lorentz symmetry breaking, Neutrino oscillations.}

\maketitle

\pagestyle{myheadings}


\section{Introduction}
The Standard Model (SM) is the most successful framework to describe the interaction among the elementary particles at the electroweak (EW) scale. However, many experimental features indicate the SM as an effective theory and it must be part of a more fundamental theory.
The neutrino's oscillation phenomena measure the squared difference of the neutrinos masses and indicate that the SM needs to be extended to include masses for the neutrinos \cite{Araki}.
The difference squared of neutrino's masses is associated with the observed probability transition $\nu_{\mu} \rightarrow \nu_{e}$ reported by the LSND experiment \cite{AthanassopoulosPRL96,AthanassopoulosPRL98}.

Several mechanisms to generate masses to the neutrinos are known in the literature \cite{seesaw1,seesaw2,SenjanovicPRD,seesaw3,seesaw4,seesaw5}.
The most famous is the seesaw mechanism in which right-handed neutrinos are introduced in models beyond the SM to provide
the largest particle content with a rich phenomenology that can be detected in accelerators in the future. In these extended models, the scalar sector after the spontaneous symmetry breaking (SSB) yields masses to the left-handed neutrinos (light mass), and the right-handed neutrinos (heavy mass). This is known as a type-II seesaw mechanism. Some phenomenological models involving Lorentz symmetry breaking have been proposed as well \cite{kost1,Katori, DIAS}.

In the context of Majorana particles, is well known that the possibilities for built fermionic bilinears are constrained due to Majorana conditions. For instance, with one Majorana spinor, there are only three possibilities, which are the scalar, pseudoscalar, and pseudo-vectorial (axial) bilinears \cite{ChengLi}.
One can use these bilinears to form quartic interactions, and through both perturbative and non-perturbative approaches, one can show that these models are a rich environment to dynamical symmetry breaking (DSB) occurs \cite{Rosenstein, Higashijima,VictoriaPRL,Miranski,NJL}. Interestingly, one can analyze one special four-fermion interaction for Majorana particles - the axial one. As can be seen in refs. \cite{DLSB1,DLSB2,DLSB3,DLSB4,DLSB5,YuriPRD2021} the axial four-fermion interaction can dynamically generate the breaking of the Lorentz symmetry by the non-null vacuum expected value (VEV) of the axial bilinear $\langle \overline{\psi} \gamma^\mu \gamma_5 \psi \rangle \neq 0$, and therefore the vacuum state of the model violates the Lorentz symmetry, and also the CP and CPT symmetries.

In this paper, we propose a model with two Majorana fermions that couple through the quartic interactions. After a type-II seesaw mechanism, in the physical field basis, one Majorana fermion acquires a light mass (that can describe a light neutrino), while the other one has a heavy mass that can be fixed at a high energy scale. On a physical basis, the Majorana fermions have two axial quartic self-coupling, and the third coupling involves the axial current of the light fermion contracted with the axial current of the heavy-fermion at the tree level. This model is an extension of the 4D Thirring model with a light fermion and another heavy-fermion that couple between itself through quartic interactions. However, this model is not renormalizable. It must be thought of as a low-energy effective theory and as part of a more fundamental theory. In this sense, the proposal is analogous to the Nambu-Jona-Lasinio (NJL) model for QCD \cite{NJL}.

We introduce two auxiliaries gauge fields to obtain the effective potential of the model. The ground state of the model is obtained in terms of two constant $4$-vectors where the effective potential is minimized. We examine the behavior of the model around these two vacuum expected values (VEVs) to generate radiative contributions to the action.  We calculate the radiative corrections up to second-order to obtain the propagation terms for the auxiliaries gauge bosons. The massive terms generated are mixed and depend on the Majorana masses from the seesaw mechanism: one gauge field has a light mass, while another one has a heavy mass. Since the mixing is very weak, a term Chern-Simons term appears associated with the heavy gauge field. As a consequence of these radiative corrections, the Lorentz and CPT symmetry is broken spontaneously. Therefore, we obtain the correspondent dispersion relations for the Majorana fermions and gauge bosons comparing them with the results already known in the literature. As an application of the seesaw mechanism, we also examine the neutrino oscillations calculating the transition probability of $\nu_{e} \rightarrow \nu_{\mu}$, and the oscillation length in the strong quartic coupling
limit. We discuss the influence of the DLSB in the parameter space of this transition.

The paper is organized as follows. In section \ref{sec2}, we review the type-II seesaw mechanism and propose the most general quartic
couplings for two Majorana fermions. Section \ref{sec3} focuses on the effective model in which are introduced two auxiliaries gauge fields, and we
obtain the effective potential and the dispersion relations for the Majorana fermions with an LSV scenario. In section \ref{sec4},
we calculate the effective Lagrangian up to second-order and the dispersion relations for the gauge bosons induced by the field fluctuations
around the ground state of the model. Section \ref{sec5} is dedicated to the application of the neutrino's oscillations
in which we examine the probability in the transition $\nu_{e} \rightarrow \nu_{\mu}$ and we obtain the parameter space.
Our conclusions and final remarks are cast in Section \ref{sec6}.
We adopt the convention for the metric $\eta^{\mu\nu}=\mbox{diag}\left(+1,-1,-1,-1\right)$, and we work with the natural units: $\hbar=c=1$.
\section{The description of the four-Majorana model}
\label{sec2}
In many models with Majorana neutrinos in the literature, the mass sector with left-handed neutrinos $(\nu_{L})$ and right-handed neutrinos
$(N_{R})$ is given by
\begin{equation}\label{Lmass}
-{\cal L}_{mass} = M \, \overline{\nu}_{L} \, N_{R} + M_{L} \, \overline{\nu_{L}^{c}} \, \nu_{L} + M_{R} \, \overline{N_{R}^{c}} \, N_{R} + \mbox{h. c.} \; ,
\end{equation}
where $M$ are $3\times 3$ matrix elements of Dirac mass term, $M_{L}$ are $3 \times 3$ matrix elements of a Majorana mass term for LHNs, and $M_{R}$ is the correspondent one to RHNs. The $\nu_{L}$ sets a column of three LHNs, {\it i. e.}, $\left\{ \, \nu_{eL} \, , \, \nu_{\mu L} \, , \, \nu_{\tau L} \, \right\}$, and $N_{R}$ is the similar column vector for the RHNs. For a brief review, a four-component Majorana fermion is defined as a
spinor $\psi$ which obeys the identity $\psi \equiv \psi^{c} = i \gamma^2 \psi^{\ast}$. The counterpart $\overline{\psi^{c}}$ is given by $\overline{\psi^{c}} = \psi^{t} \, C$, where $C=i \, \gamma^0 \, \gamma^2$ is the unitary charge conjugation matrix $C^\dagger = C^{-1}$,
$\psi^{t}$ means the transpose column matrix for the $\psi$-spinor, and $\left\{ \, \gamma^{0} \, , \, \gamma^{2} \, \right\}$ are two Dirac matrices.
Backing to the mass sector (\ref{Lmass}), it is defined the variables $\eta:=\nu_{L}+\nu_{L}^{c}$ and $\chi:=N_{R}+N_{R}^{c}$, the massive lagrangian can be written in terms of the two four-component Majorana spinors $\eta$ and $\chi$ :
%
%
\begin{eqnarray}
-\mathcal{L}_{mass} \!\!&=&\!\!
\frac{M}{2} \left( \, \overline{\eta} \, \chi + \overline{\chi} \, \eta \, \right) + M_{L} \, \overline{\eta} \, \eta + M_{R} \, \overline{\chi} \, \chi
\nonumber \\
\!\!&=&\!\!
\left(
\begin{array}{cc}
\overline{\eta} & \;\;\;
 \overline{\chi}
\\
\end{array}
\right)
\left[
\begin{array}{cc}
M_{L} & M/2
\\
\\
M/2 & M_{R} \\
\end{array}
\right]
\left(
\begin{array}{c}
\eta \\
\\
\chi \\
\end{array}
\right)
\, .
\end{eqnarray}
The mass eigenstates, that we denote by $\psi_{i}\,(i=1,2)$, are obtained by the diagonalization of the mass matrix :
\begin{subequations}
\begin{eqnarray}
\psi_1 &=& \cos \theta \, \eta - \sin \theta \, \chi \; ,
\label{transfpsi1}
\\
\psi_2 &=& \sin \theta \, \eta + \cos \theta \, \chi \; ,
\label{transfpsi2}
\end{eqnarray}
\end{subequations}
where $\tan 2\theta = \frac{M}{M_{L} - M_{R}}$, and the correspondent eigenvalues are given by :
\begin{subequations}
\begin{eqnarray}
m_{1} \!=\! \frac{1}{2}\left(M_{L} + M_{R}\right) - \frac{1}{2} \sqrt{(M_{L}-M_{R})^2 + M^2} \; ,
\;\;\;
\\
m_{2} \!=\! \frac{1}{2}\left(M_{L} + M_{R}\right) + \frac{1}{2} \sqrt{(M_{L}-M_{R})^2 + M^2} \; . \;\;\;
\end{eqnarray}
\end{subequations}
The condition $M_{R} \gg M_{L} \gg M$ is applied for the neutrino's case. Under this consideration,
the previous eigenvalues are read below :
\begin{subequations}
\begin{eqnarray}
m_{1} &\simeq& M_{L}-\frac{M^2}{4M_{R}} \simeq M_{L} \; ,
\\
m_{2} &\simeq& M_{R}+\frac{M^2}{4M_{R}}\simeq M_{R} \; ,
\end{eqnarray}
\end{subequations}
and the mixing angle is very small $\theta\simeq -M/(2M_{R})\ll 1$, where we can write $\cos\theta \simeq 1$ and $\sin\theta\simeq0$ in
(\ref{transfpsi1}) and (\ref{transfpsi2}). Thereby, $\psi_{1}$ is identified as a light Majorana fermion (left-handed neutrinos),
and $\psi_{2}$ is a heavy Majorana fermion that describes a right-handed neutrino with mass defined on the TeV scale or higher.
For one Majorana $\psi$-spinor, the non-trivial bilinears which can be formed are :
%
$\overline{\psi} \, \psi \, , \,
\overline{\psi} \, \gamma_5 \, \psi
\, , \,
\overline{\psi} \, \gamma^\mu \, \gamma_5 \, \psi$.
%
Otherwise, any bilinear combination is null.
%
%
%
%
%
%
Using the properties of the Majorana spinors, the non-trivial bilinear involving $\eta$ and $\chi$ are the combinations below :
\begin{eqnarray}
\overline{\eta} \, \gamma^\mu \, \gamma_5 \, \eta
\; , \;
\overline{\chi} \, \gamma^\mu \, \gamma_5 \, \chi
\; , \;
\overline{\chi} \, \gamma^\mu \, \gamma_5 \, \eta + \overline{\eta} \, \gamma^\mu \, \gamma_5 \, \chi
\; , \;
\nonumber \\
\overline{\eta} \, \gamma^\mu \, \chi - \overline{\chi} \, \gamma^\mu \, \eta \; .
\end{eqnarray}
Following these combinations, we propose the most general pseudo-vector quartic self-interaction for the Majorana
fermions $\eta$ and $\chi$  :
\begin{eqnarray}\label{Lint}
\mathcal{L}^{int} \!&=&\!  \frac{G_1}{2} \left( \, \overline{\eta} \, \gamma_\mu \, \gamma_5 \, \eta \, \right)^2
+ \frac{G_2}{2} \left(\overline{\chi} \, \gamma_\mu \, \gamma_5 \, \chi \right)^2
\nonumber \\
&&
\hspace{-0.3cm}
+ \frac{G_3}{2} \left( \overline{\eta} \, \gamma_\mu \, \gamma_5 \, \chi + \overline{\eta} \, \gamma_\mu \, \gamma_5 \, \chi \right)^2
\nonumber \\
&&
\hspace{-0.3cm}
+ \frac{G_4}{2} \left( \overline{\eta} \, \gamma_\mu \, \chi - \overline{\chi} \, \gamma_\mu \, \eta \right)^2 \; ,
\end{eqnarray}
where $G_{i} \, (i=1,2,3,4)$ are coupling constants with length dimension squared. Using the Fierz identities
\begin{subequations}
\begin{eqnarray}
&&
\frac{1}{4} \left( \, \overline{\eta} \, \gamma_\mu \, \chi - \overline{\chi} \, \gamma_\mu \, \eta\right)^2 = \left(\overline{\eta} \, \eta\right)\left(\overline{\chi} \, \chi\right)
\nonumber \\
&&
-  \left(\overline{\eta} \, \gamma_5 \, \eta\right)\left(\overline{\chi} \, \gamma_5 \, \chi\right)
- \frac{1}{2} \left(\overline{\eta} \, \gamma_\mu \, \gamma_5 \, \eta\right)\left(\overline{\chi} \, \gamma^\mu \, \gamma_5 \, \chi\right)
\, , \hspace{0.6cm}
\\
&&
\frac{1}{4} \left(\overline{\eta} \, \gamma_\mu \, \gamma_5 \, \chi + \overline{\eta} \, \gamma_\mu \, \gamma_5 \, \chi\right)^2 = -\left(\overline{\eta} \, \eta\right)\left( \, \overline{\chi} \, \chi \, \right)
\nonumber \\
&&
+  \left(\overline{\eta} \, \gamma_5 \, \eta\right)\left(\overline{\chi} \, \gamma_5 \, \chi\right)
- \frac{1}{2} \left(\overline{\eta} \, \gamma_\mu \, \gamma_5 \, \eta\right)\left(\overline{\chi} \, \gamma^\mu  \, \gamma_5 \, \chi\right)
\, , \hspace{0.6cm}
\end{eqnarray}
\end{subequations}
the couplings from (\ref{Lint}) are written in the physical eigenstates basis $\left\{ \, \psi_1 \, , \, \psi_2 \, \right\}$ as :
\begin{eqnarray}
\mathcal{L}^{int} \!&=&\!  -\frac{G_1}{2} \, ( \overline{\psi}_1 \, \gamma_\mu \, \gamma_5 \, \psi_1 )^2
- \frac{G_2}{2} \, ( \overline{\psi}_2 \, \gamma_\mu \, \gamma_5 \, \psi_2 )^2
\nonumber \\
&&
\hspace{-1.5cm}
+ \frac{1}{2}\left(G_{3}+G_{4}\right) \, (\overline{\psi}_1 \, \gamma_\mu \, \gamma_5 \, \psi_1)(\overline{\psi}_2 \, \gamma^\mu \, \gamma_5 \, \psi_2)
\nonumber \\
&&
\hspace{-1.5cm}
- (G_{3}-G_{4})\left[ (\overline{\psi}_{1}\psi_{1})(\overline{\psi}_{2}\psi_{2})\!-\!(\overline{\psi}_{1}\gamma_{5}\psi_{1})(\overline{\psi}_{2}\gamma_{5}\psi_{2}) \right]
. \; \; \;\; \;
\end{eqnarray}
where we have considered the small mixing angle. We choose the case in which $G_{3}=G_{4}$, such that the NJL-like terms that could contribute with the
dynamically generated masses can be eliminated, and we will focus just in the terms with the axial currents of $\psi_{1}$ and $\psi_{2}$. In this particular case, the couplings are read :
\begin{eqnarray}
\mathcal{L}^{int} \!\!&=&\!\!  -\frac{G_1}{2} \, ( \overline{\psi}_1 \, \gamma_\mu \, \gamma_5 \, \psi_1 )^2
- \frac{G_2}{2} \, (\overline{\psi}_2 \, \gamma_\mu \, \gamma_5 \, \psi_2)^2
\nonumber \\
&&
\hspace{-0.5cm}
+ \, G_{3} \, (\overline{\psi}_1 \, \gamma_\mu \, \gamma_5 \, \psi_1)(\overline{\psi}_2 \, \gamma^\mu \, \gamma_5 \, \psi_2) \; .
\end{eqnarray}
%
%
%
%
%
%
%
%
Therefore, the model contains axial couplings for description of the processes $\psi_{1} \, \psi_{1} \, \rightarrow \, \psi_{1} \, \psi_{1}$, $\psi_{2} \, \psi_{2} \, \rightarrow \, \psi_{2} \, \psi_{2}$, and $\psi_{1} \, \psi_{1} \, \rightarrow \, \psi_{2} \, \psi_{2}$ at tree level. This is an extension of the model studied in ref. \cite{DLSB1}, in which we will investigate the dynamical symmetry breaking.

\section{The effective potential and the dispersion relations for the Majorana fermions}
\label{sec3}
\begin{figure*}[th]
\centering
\includegraphics[width=0.325\textwidth]{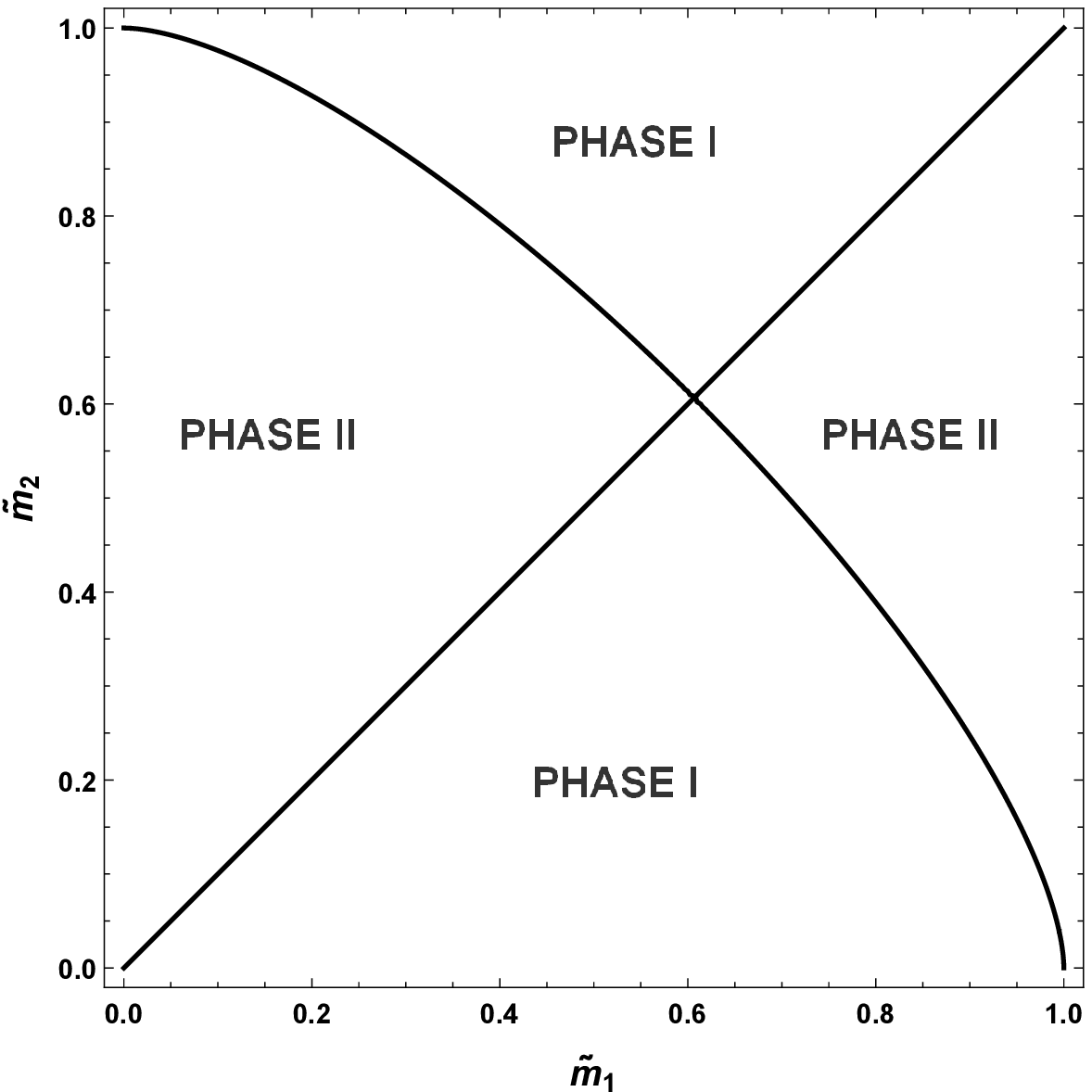}
\includegraphics[width=0.325\textwidth]{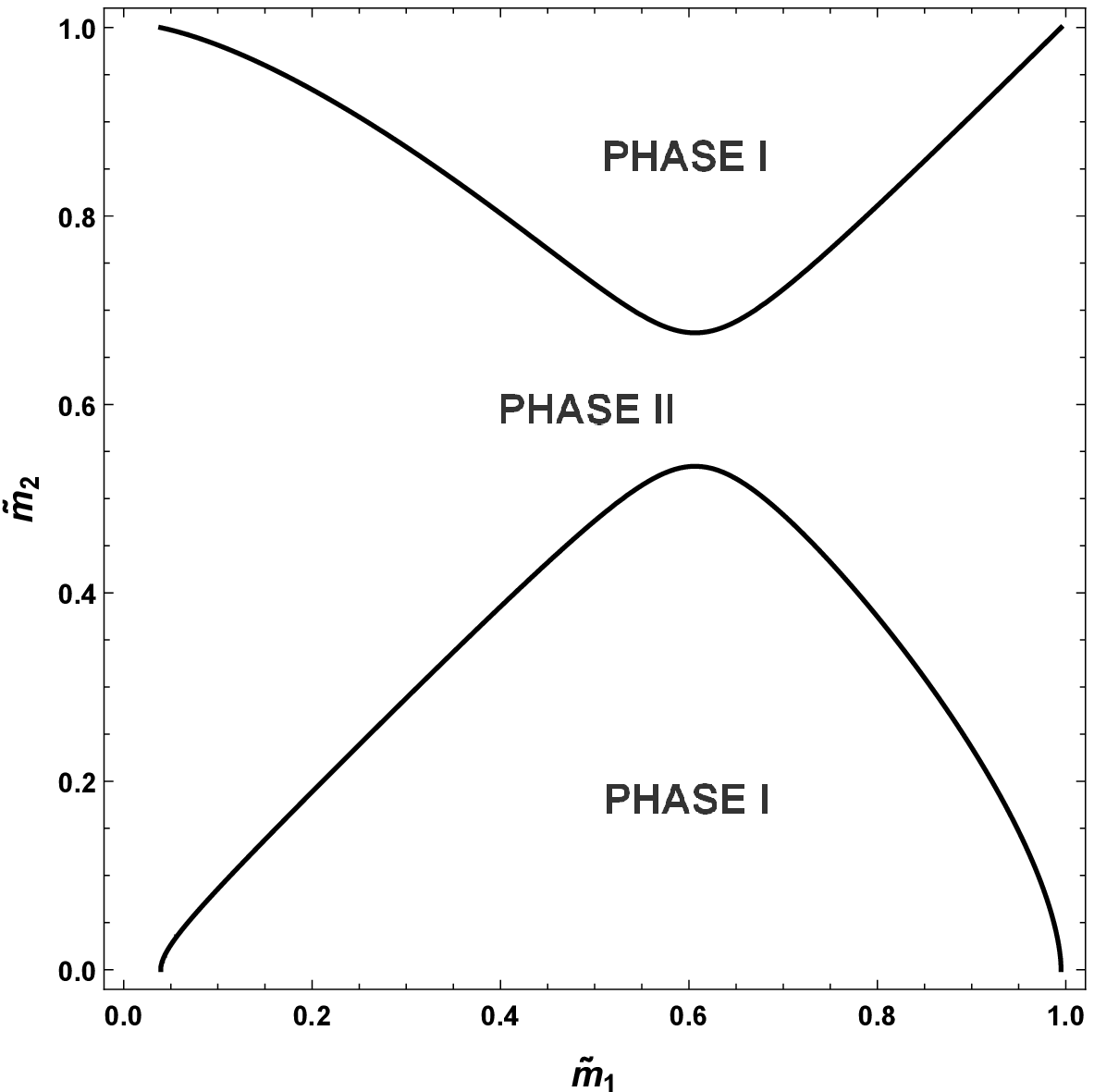}
\includegraphics[width=0.325\textwidth]{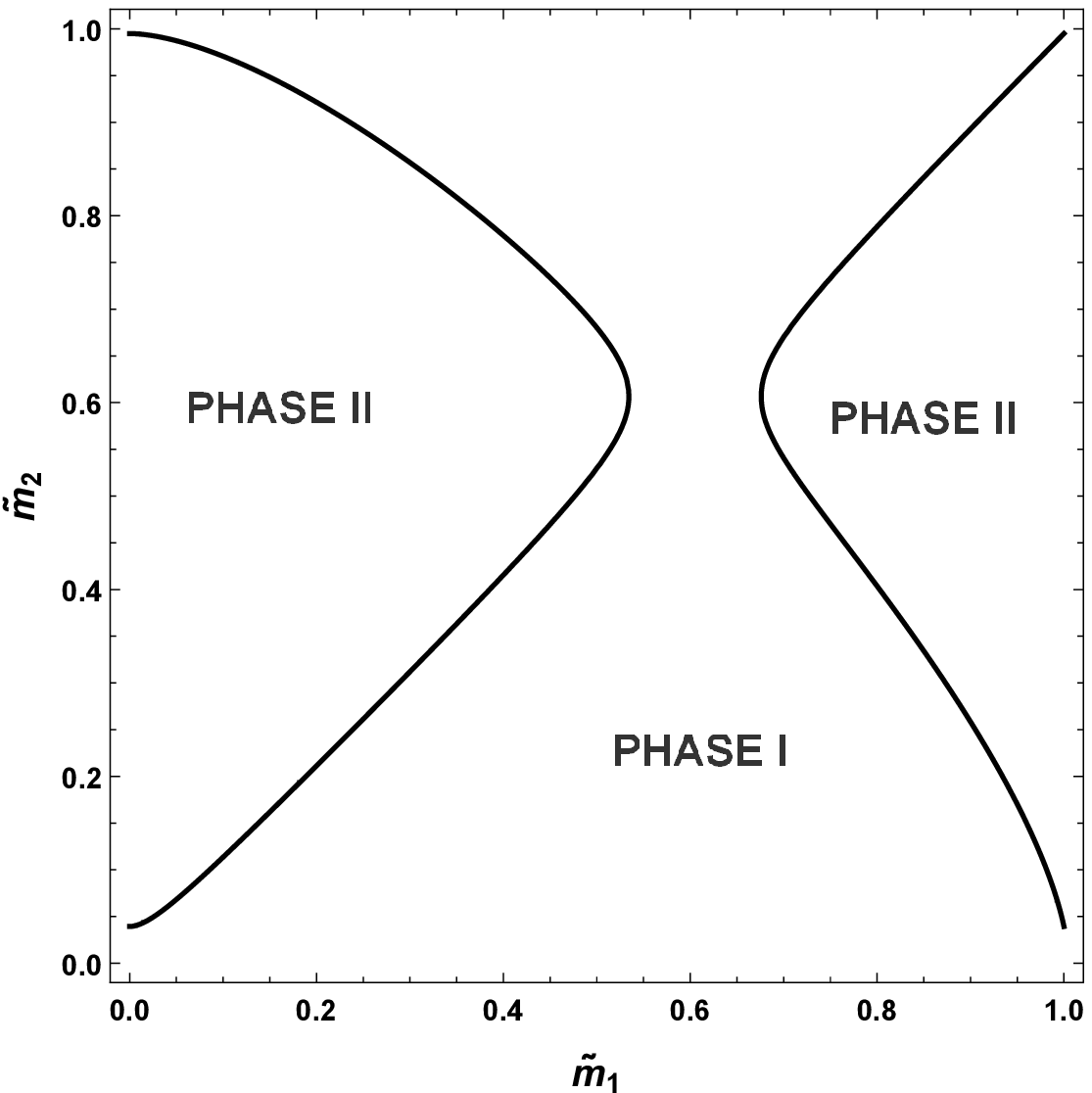}
\caption{The phase diagram of the model for $G_3 \neq 0$, with the conditions $G_2>G_1$ (left panel), $G_1=G_2$ (middle panel)
and $G_1>G_2$ (right panel), respectively.
The phase I refers to the region in which $\beta^{\mu} = 0$ and $\alpha^{\mu} \neq 0$, while the phase II region correspond to $\alpha^{\mu} = 0$ and $\beta^{\mu} \neq 0$. The masses are normalized in energy scale unit.}
\label{phasespace}
\end{figure*}

We consider in this section the model with the two Majorana fermions and their quartic self-interactions
through the axial currents :
\begin{eqnarray}
{\cal L}_{model} \!\!&=&\!\!
\overline{\psi}_{1} ( \, i \, \slash{\!\!\!\partial} - m_{1} \, )\psi_{1}
-\frac{G_1}{2} \, ( \overline{\psi}_1 \, \gamma_\mu \, \gamma_5 \, \psi_1)^2
\nonumber \\
&&
\hspace{-0.7cm}
+\,\overline{\psi}_{2}( \, i \, \slash{\!\!\!\partial} - m_{2} \, )\psi_{2}
-\frac{G_2}{2} \, (\overline{\psi}_2 \, \gamma_\mu \, \gamma_5 \, \psi_2)^2
\nonumber \\
&&
\hspace{-0.7cm}
+\,G_3 \, (\overline{\psi}_1 \, \gamma_{\mu} \, \gamma_5 \, \psi_1)
(\overline{\psi}_2 \, \gamma^{\mu} \, \gamma_5 \, \psi_2 ) \; .
\end{eqnarray}
This Lagrangian is equivalent to
\begin{eqnarray}
{\cal L}_{model} \!&=&\! \overline{\psi}_{1}\left( \, i \, \slash{\!\!\!\partial} -g \, \, \slash{\!\!\!\!A} \, \gamma_{5} - m_{1} \, \right)\psi_{1}
\nonumber \\
&&
\hspace{-0.8cm}
+ \, \overline{\psi}_{2}\left( \, i \, \slash{\!\!\!\partial}-g^{\prime} \,\, \slash{\!\!\!\!B} \, \gamma_{5} - m_{2} \, \right)\psi_{2}
\nonumber \\
&&
\hspace{-0.8cm}
+ \, \frac{1}{2}\, g_{1}^2 \,A_{\mu}A^{\mu}+\frac{1}{2}\, g_{2}^2 \,B_{\mu}B^{\mu}+g_{3}^2 \, A_{\mu}B^{\mu} \, ,
\hspace{0.5cm}
\end{eqnarray}
where the auxiliary fields $A^{\mu}$ and $B^{\mu}$, using the motion equations, satisfy the constraints
\begin{subequations}
\begin{eqnarray}
A^{\mu}\!=\!\frac{g_{2}^2}{g_{1}^2g_{2}^2-g_{3}^4}\!\left[ g \, \overline{\psi}_{1} \gamma^{\mu}\gamma_{5}\psi_{1}
-\frac{g_3^2}{g_2^2} \, g^{\prime} \, \overline{\psi}_{2} \gamma^{\mu}\gamma_{5}\psi_{2} \right] ,
\nonumber \\
\\
B^{\mu}\!=\!\frac{g_{1}^2}{g_{1}^2g_{2}^2-g_{3}^4}\!\left[ g^{\prime} \, \overline{\psi}_{2} \gamma^{\mu}\gamma_{5}\psi_{2}
-\frac{g_3^2}{g_1^2} \, g \, \overline{\psi}_{1} \gamma^{\mu}\gamma_{5}\psi_{1} \right] ,
\nonumber \\
\end{eqnarray}
\end{subequations}
and the coupling constants $G_{i} \, (i=1,2,3)$ are parameterized by
\begin{eqnarray}
G_{1}:=\frac{g^2 \, g_{2}^2}{g_{1}^{2}g_{2}^2-g_{3}^4}
\hspace{0.3cm} , \hspace{0.3cm}
G_{2}:=\frac{g^2 \, g_{1}^2}{g_{1}^{2}g_{2}^2-g_{3}^4}
\nonumber \\
\hspace{0.3cm} \mbox{and} \hspace{0.3cm}
G_{3}:=\frac{ g \, g^{\prime} \, g_{3}^2}{g_{1}^{2}g_{2}^2-g_{3}^{4}} \; .
\end{eqnarray}
Notice also that the constants $g_{i} \, (i=1,2,3)$ have mass dimension, while that
$g$ and $g^{\prime}$ are dimensionless coupling constants of the fermions $\psi_{1}$ and $\psi_{2}$,
with the auxiliary gauge fields $A^{\mu}$ and $B^{\mu}$, respectively. The perturbative formalism allow
us to define the functional integration
\begin{eqnarray}
Z=\int {\cal D}A^{\mu}{\cal D}B^{\mu}{\cal D}\overline{\psi}_{1}{\cal D}\psi_{1}{\cal D}\overline{\psi}_{2}{\cal D}\psi_{2} \, e^{i \int d^4x \, {\cal L}_{model}}
\nonumber \\
= \int {\cal D}A^{\mu}{\cal D}B^{\mu} \, e^{i S_{eff}(A,B)} \; ,
\hspace{0.4cm}
\end{eqnarray}
where, after the fermion integrations, we obtain the effective action
\begin{eqnarray}
S_{eff}(A,B) \!\! &=& \!\!\! \int \! d^4x \! \left[ \frac{1}{2}\, g_1^2 \, A^{2}+\frac{1}{2}\, g_{2}^2 \, B^{2}+ g_{3}^2 \, A \cdot B \right]
\nonumber \\
&&
-i \, \mbox{Tr}\ln\left( i \, \slash{\!\!\!\partial} - g \, \slash{\!\!\!\!A} \gamma_{5} - m_1 \right)
\nonumber \\
&&
-i \, \mbox{Tr}\ln\left( i \, \slash{\!\!\!\partial - g^{\prime} \, \slash{\!\!\!\!B} \gamma_{5}} - m_2 \right) \; ,
\end{eqnarray}
and Tr means the trace on the Dirac matrices and on the coordinate, or on the momentum space. The correspondent effective potential is
\begin{eqnarray}
V_{eff}(A,B) \!\!&=&\!\! -\frac{1}{2}\, g_{1}^2 \,A^{2}
-\frac{1}{2}\, g_{2}^2 \, B^{2}-g_{3}^2 \, A\cdot B
\nonumber \\
&&
\hspace{-1.3cm}
+\, i \int \frac{d^4p}{(2\pi)^4} \, \mbox{tr}\left[ \ln\left( \, \slash{\!\!\!p} - m_1 - g \, \slash{\!\!\!\!A} \gamma_{5} \, \right) \right]
\nonumber \\
&&
\hspace{-1.3cm}
+ \, i \int \frac{d^4p}{(2\pi)^4} \, \mbox{tr}\left[ \ln\left( \, \slash{\!\!\!p} - m_2 - g^{\prime} \, \slash{\!\!\!\!B} \gamma_{5}  \, \right) \right]
 .
\hspace{0.6cm}
\end{eqnarray}
This potential has two non-trivial minimal :
\begin{subequations}
\begin{eqnarray}
\left.\frac{\partial V_{eff}}{\partial A_{\mu}}\right|_{A=\alpha,B=\beta} \!\!\!&=&\!\!\! -g_{1}^2\,\alpha^{\mu}- g_3^2 \, \beta^{\mu}
+ i \, \Pi_{1}^{\mu}   = 0 \, ,
\hspace{0.8cm}
\\
&&
\hspace{0.3cm} \mbox{and} \hspace{0.3cm}
\label{partialA}
\nonumber \\
\left.\frac{\partial V_{eff}}{\partial B_{\mu}}\right|_{A=\alpha,B=\beta} \!\!\!&=&\!\!\! -g_{2}^2\,\beta^{\mu}-g_3^2 \, \alpha^{\mu}
+ i \, \Pi_{2}^{\mu}  = 0 \, ,
\label{partialB}
\hspace{0.8cm}
\end{eqnarray}
\end{subequations}
where
\begin{subequations}
\begin{eqnarray}
\Pi_{1}^{\mu}= \int \frac{d^4p}{(2\pi)^4} \, \mbox{tr}\left[\frac{1}{\slash{\!\!\!p} - m_1 - \slash{\!\!\!a} \, \gamma_{5}}
\, (-g) \gamma^{\mu}\gamma_{5}\right] \; ,
\hspace{0.5cm}
\\
\Pi_{2}^{\mu}=\int \frac{d^4p}{(2\pi)^4} \, \mbox{tr}\left[\frac{1}{\slash{\!\!\!p} - m_2 - \slash{\!\!\!b} \, \gamma_{5}} \, (-g^{\prime}) \gamma^{\mu}\gamma_{5}\right]  \; ,
\hspace{0.5cm}
\end{eqnarray}
\end{subequations}
and we have defined $a^{\mu}=g\alpha^{\mu}$, e $b^{\mu}=g^{\prime}\beta^{\mu}$. Both integrals are like the tadpole diagrams and diverge in the ultraviolet limit. We introduce the dimensional regularization (with a $D$ regulator parameter) to calculate these integrals, and consequently, we can isolate the divergent terms. The coupling constants $g$ and $g^{\prime}$ are replaced by $g \rightarrow g \, (\mu^2)^{1-d/2}$ and $g^{\prime} \rightarrow g^{\prime} \, (\mu^2)^{1-d/2}$, where $\mu$ is an arbitrary energy scale. After some manipulations, the trace calculus yields the regularized integrals
\begin{eqnarray}
&&
\Pi_{1}^{\mu}(D)=-4g\left(\mu^2\right)^{1-d/2} \int \frac{d^{D}p}{(2\pi)^{D}} \times
\nonumber \\
&&
\hspace{-0.5cm}
\times \,\frac{2[-p^2a^{\mu}+(p\cdot a)p^{\mu}]+(p^2-m_{1}^2-a^2)a^{\mu}}{(p^2-m_{1}^2-a^2)^{2}+4[p^2a^2-(p\cdot a)^{2}]}
\, , \hspace{0.5cm}
\\
&&
\Pi_{2}^{\mu}(D) = -4g^{\prime}\left(\mu^2\right)^{1-d/2} \int \frac{d^{D}p}{(2\pi)^{D}} \times
\nonumber \\
&&
\hspace{-0.5cm}
\times \, \frac{2[-p^2b^{\mu}+(p\cdot b)p^{\mu}]+(p^2-m_{2}^2-b^2)b^{\mu}}{(p^2-m_{2}^2-b^2)^{2}+4[p^2b^2-(p\cdot b)^{2}]} \, .
\end{eqnarray}
Using the approximation $\left\{ \, m_{1}^2 \, , \, m_{2}^2 \, \right\} \gg \left\{ \, a^2 \, , \, b^2 \, \right\}$, we expand the previous integral into the $a$ and $b$-parameters to obtain the results
\begin{subequations}
\begin{eqnarray}
i\Pi_1^\mu \!\!&=&\!\! g \, a^\mu \left[  \frac{ m_1^2 }{\pi^2 \epsilon} - \frac{m_1^2 }{\pi^2 } \ln \left(\frac{m_1}{\Lambda} \right)+ \frac{a^2}{3 \pi^2} \right]
\, , \hspace{0.5cm}
\\
i\Pi_2^\mu \!\!&=&\!\! g^{\prime} \, b^\mu \left[  \frac{ m_2^2 }{\pi^2 \epsilon} - \frac{m_2^2 }{\pi^2 } \ln \left(\frac{m_2}{\Lambda} \right)+ \frac{b^2}{3 \pi^2} \right] \, , \hspace{0.7cm}
\end{eqnarray}
\end{subequations}
where we have used the physical dimension in $D=4-\epsilon$, and $\Lambda^2:=4\pi\mu^{2} \, e^{-\gamma}$, in which $\gamma=0.577$ is the Euler-Mascheroni constant. Therefore, we substitute these results in (\ref{partialA}) and (\ref{partialB}), respectively, and we obtain the relations :
\begin{subequations}
\begin{eqnarray}
\frac{g_{3}^2}{gg^{\prime}}\,b^{\mu}=
a^\mu \! \left[-\frac{g_{1}^2}{g^2}+  \frac{ m_1^2 }{\pi^2 \epsilon} - \frac{m_1^2 }{\pi^2 } \ln \left(\frac{m_1}{\Lambda} \right)+ \frac{a^2}{3 \pi^2} \right] \; ,
\label{gapa}
\hspace{-0.5cm}
\nonumber \\
\\
\frac{g_3^2}{gg^{\prime}}\,a^{\mu}
= b^\mu \! \left[-\frac{g_2^2}{g^{\prime2}}+  \frac{ m_2^2 }{\pi^2 \epsilon} - \frac{m_2^2 }{ \pi^2 } \ln \left(\frac{m_2}{\Lambda} \right)+ \frac{b^2}{3 \pi^2} \right] \, .
\label{gapb}
\hspace{-0.5cm}
\nonumber \\
\end{eqnarray}
\end{subequations}
From the above equations, we have two non-trivial solutions :
\begin{enumerate}
\item Phase I - The case of $b^{\mu}=0$, the solution for $a^{\mu} \neq 0$ is given by
\begin{eqnarray}
-\frac{g_{1R}^2}{g^2}- \frac{m_1^2 }{\pi^2 } \ln \left(\frac{m_1}{\Lambda} \right)+ \frac{a^2}{3 \pi^2}=0 \; .
\end{eqnarray}
\item Phase II - For the case of $a^\mu = 0$ and $b^{\mu} \neq 0$, the solution is
\begin{eqnarray}
-\frac{g_{2R}^2}{g^{\prime2}}- \frac{m_2^2}{\pi^2 } \ln \left(\frac{m_2}{\Lambda} \right)+ \frac{b^2}{3 \pi^2}=0 \; .
\end{eqnarray}
\end{enumerate}
The phase space is illustrated in the figure (\ref{phasespace}) including the three cases : $G_{2} > G_{1}$ (right panel),
$G_{2} = G_{1}$ (middle panel) and $G_{1} > G_{2}$ (right panel). We use $G_{3} \neq 0$ in all plots. The masses
$(\tilde{m}_{1},\tilde{m}_{2})$ are normalized by the energy scale $(\mu)$, {\it i.e.}, $\tilde{m}_{i}\equiv m_{i}/\mu \, (i=1,2)$.

It is important to highlight that only in the case $g_3=0$, both pseudo-vectors can acquire non-null vacuum expected values.  Notice that, in both solutions, we have defined the renormalizable coupling constants
\begin{eqnarray}
\frac{g_{1R}^2}{g^{2}} =\frac{g_{1}^2}{g^2} -\frac{ m_1^2 }{\pi^2 \epsilon}
\hspace{0.3cm} \mbox{and} \hspace{0.3cm}
\frac{g_{2R}^2}{g^{\prime2}} =\frac{g_{2}^2}{g^{\prime2}} -\frac{ m_2^2 }{\pi^2 \epsilon} \; .
\end{eqnarray}

The true solution of the gap equation is defined by the global minimum of the potential. Integrating the gap equations,
we obtain the effective potential:
\begin{eqnarray}\label{poteff}
V_{eff}(A,B) = \frac{g^2}{12\pi^2} \left(A^2-a^2\right)^{2}+
\nonumber \\
+ \, \frac{g^{\prime\,2}}{12\pi^2} \left(B^2-b^{2}\right)^2-g_{3}^{2} \, A \cdot B \; .
\end{eqnarray}
Thereby, true minimum of the potential is at the points
\begin{subequations}
\begin{eqnarray}
a^2 \!\!&=&\!\! 3\pi^2 \left[ \frac{g_{1R}^2}{g^{2}}+ \frac{m_1^2 }{\pi^2 } \ln \left(\frac{m_1}{\Lambda} \right) \right]
\hspace{0.15cm} \mbox{and} \hspace{0.15cm}
b^\mu=0 \; ,
\nonumber \\
\\
b^2 \!\!&=&\!\! 3\pi^2 \left[\frac{g_{2R}^2}{g^{\prime2}}+ \frac{m_2^2 }{\pi^2 } \ln \left(\frac{m_2}{\Lambda} \right)\right]
\hspace{0.15cm} \mbox{and} \hspace{0.15cm}
a^\mu=0 \; .
\nonumber \\
\end{eqnarray}
\end{subequations}
If we consider a very small mass for $\psi_{1}$-majorana fermion, the first minimum point is determinate by the
renormalized coupling constant $g_{1R}$, {\it i. e.}, $a^2=3\pi^2g_{1R}^2/g^2$. The effective potential has a
nontrivial minimal in which $\langle A^{\mu} \rangle = \alpha^{\mu}$ and $\langle B^{\mu} \rangle = \beta^{\mu}$
are two scales of VEVs that break the Lorentz symmetry.
As consequence, the modified dispersion
relations for the Majorana fermions are read below.
In the phase I :
\begin{subequations}

\begin{eqnarray}
&&
(p^2-m_1^{2}-a^2)^{2}+4\left[ p^2 \, a^2 - (p\cdot a)^2 \right] = 0 \; ,
\hspace{1cm}
\label{DRm1}
\\
&&
p^2 = m_{2}^{2} \; ,
\end{eqnarray}
\end{subequations}

and in the phase II:

\begin{subequations}
\begin{eqnarray}
&&
p^2 = m_{1}^2 \; , ~~
\\
&&
(p^2-m_2^{2}-b^2)^{2}+4\left[ p^2 \, b^2 - (p\cdot b)^2 \right] = 0 \; .
\label{DRm2}
\hspace{1cm}
\end{eqnarray}
\end{subequations}
These dispersion relations are analogous to Carroll-Field-Jackiw-Proca electrodynamics \cite{CFJED}.
The frequency solutions from (\ref{DRm1}) and (\ref{DRm2}) are hard to obtain
in this present form. Thus, we can consider the particular cases of a time-like and space-like DLSB parameter.
For a time-like case, in which $a^{\mu}=(a^{0},{\bf 0})$ and $b^{\mu}=(b^{0},{\bf 0})$, the frequency solutions are read below :
%
\begin{eqnarray}\label{omega1m1peq}
\omega_{1}^{(\pm)}({\bf p}) \!\!&=&\!\! \sqrt{(|{\bf p}| \pm a_0)^{2}+m_{1}^2}
\nonumber \\
\!\!&\simeq&\!\! ||{\bf p}| \pm a_0|+\frac{m_{1}^2}{2||{\bf p}| \pm a_0|}
 \; ,
\end{eqnarray}
%
and
%
\begin{eqnarray}
\omega_{2}^{(\pm)}({\bf p}) \!\!&=&\!\! \sqrt{(|{\bf p}|\pm b_0)^{2}+m_{2}^2}
\nonumber \\
\!\!&\simeq&\!\! m_{2}+\frac{(|{\bf p}| \pm b_0)^2}{2m_{2}}
 \; .
\end{eqnarray}
%
For the space-like case, where $a^{\mu}=(0,{\bf a})$ and $b^{\mu}=(0,{\bf b})$,
the frequency solutions are given by
%
\begin{eqnarray}
\omega_{1}^{(\pm)}({\bf p}) \!\!&=&\!\! \sqrt{{\bf p}^2+m_{1}^2+{\bf a}^2\pm2|{\bf a}|\sqrt{m_{1}^2+({\bf p}\cdot\hat{{\bf a}})^{2}}}
\nonumber \\
\!\!&\simeq&\!\! |{\bf p}\pm{\bf a}|+\frac{m_{1}^2}{2|{\bf p}\pm{\bf a}|}
 \; ,
\end{eqnarray}
%
and
%
\begin{eqnarray}
\omega_{2}^{(\pm)}({\bf p}) \!\!&=&\!\! \sqrt{{\bf p}^2+m_{2}^2+{\bf b}^2\pm2|{\bf b}|\sqrt{m_{2}^2+({\bf p}\cdot\hat{{\bf b}})^{2}}}
\nonumber \\
\!\!&\simeq&\!\! m_{2}\pm|{\bf b}|+\frac{{\bf p}^2}{2m_{2}}
 \; .
\end{eqnarray}
%
We have considered the condition from the seesaw mechanism $m_{2}\gg m_{1}$ in the previous approximation.
For the massless case, for example, if we consider $\psi_{1}$ a Majorana fermion with a very small mass in which we can neglect it $m_{1} \approx 0$ with a generic DLSB parameter, the frequencies are given by
\begin{subequations}
\begin{eqnarray}\label{omega1omega2}
\omega_{1}^{\pm}({\bf p}) \!\! &=& \!\! -a_{0} \, \pm \, |{\bf p}+{\bf a}| \; ,
\hspace{0.8cm}
\\
\omega_{2}^{\pm}({\bf p}) \!\! &=& \!\! a_{0} \, \pm \, |{\bf p}-{\bf a}| \; .
\hspace{0.8cm}
\end{eqnarray}
\end{subequations}
The usual RDs are recovered when $a^{\mu}\rightarrow 0$. Other important point is that the results (\ref{omega1omega2}) reproduce
asymmetric frequencies and it depend on the direction of the vector ${\bf a}$ with a propagation direction $\hat{{\bf p}}$.

\section{The effective action}
\label{sec4}
The vacuum properties of the model help us to understand the dynamics of the fluctuations dictated by the auxiliary fields $A^{\mu}$
and $B^{\mu}$. Thereby, we expand these auxiliary fields around the vacuum minimal as : $A^{\mu}(x) = \alpha^\mu + X^{\mu}(x)$ and
$B^\mu(x) = \beta^\mu + Y^{\mu}(x)$, respectively, where $X^\mu$ and $Y^\mu$ are interpreted as the new
dynamical fields of the model, and $\left\{ \, \alpha^{\mu} \, , \, \beta^{\mu} \, \right\}$ do not depend on the space-time coordinates.
The effective action in terms of $X^\mu$ and $Y^\mu$ is

\begin{eqnarray}
&&
S_{eff}(X,Y) = \!\! \int d^4x \, \Bigg\{
\frac{1}{2} \, g_{1}^2 \, \alpha^2 + \frac{1}{2} \, g_{2}^{2} \, \beta^2
\nonumber \\
&&
+g_{3}^2 \, (\alpha\cdot\beta)
+X\cdot(g_{1}^2\alpha+g_{3}^2\beta)
+Y\cdot(g_{2}^2\beta+g_{3}^2\alpha)
\nonumber \\
&&
+\frac{1}{2} \, g_{1}^2 \, X^2+\frac{1}{2} \, g_{2}^{2} \, Y^2
+g_{3}^2 \, (X\cdot Y)
\nonumber \\
&&
- i \! \int \frac{d^4p}{(2\pi)^4} \, \mbox{tr}\left[ \ln\left( \, \slash{\!\!\!p} - m_1 - \slash{\!\!\!a} \, \gamma_{5}
- g \,\, \slash{\!\!\!\!X} \, \gamma_{5} \, \right) \right]
\nonumber \\
&&
- i \! \int \frac{d^4p}{(2\pi)^4} \, \mbox{tr}\left[ \ln\left( \, \slash{\!\!\!p} - m_2 - \slash{\!\!\!b} \, \gamma_{5} - g^{\prime} \, \slash{\!\!\!Y} \, \gamma_{5}  \, \right) \right] \Bigg\} .
\nonumber \\
\end{eqnarray}
Now, we expand the previous effective action in power series of $X^{\mu}$ and $Y^{\mu}$,
such that:
\begin{eqnarray}
S_{eff}(X,Y) = \sum_{n=1}^\infty S^{(n)}(X,Y) \; ,
\end{eqnarray}
where the first correction $(n=1)$ is given by
\begin{eqnarray}
S^{(1)}(X,Y) \!\!&=&\!\! i\int d^4 x \Big[ \, X_\mu \left(i\Pi_1^\mu- g_1^2 \, \alpha^\mu - g_3^2 \, \beta^\mu\right)
+
\nonumber \\
&&
\hspace{-1.2cm}
+ Y_\mu \left(i\Pi_2^\mu- g_2^2 \, \beta^\mu - g_3^2 \, \alpha^\mu\right) \, \Big]=0 \; .
\end{eqnarray}
This contribution is null, as we expect, due to the relations (\ref{partialA}) and (\ref{partialB}).
The self-energy term is the correction at the second order :
\begin{eqnarray}\label{Seff2}
S^{(2)}(X,Y) = \frac{i}{2}\sum_{i,j=1}^{2} \int d^4 x \Big[ \, V_{\mu}^{i}(x) \, \Pi^{\mu \nu}_{ij} \, V_{\nu}^{j}(x) \, \Big] \, ,
\hspace{0.3cm}
\end{eqnarray}
where we have introduced the notation $V_{\mu}^{i}= ( \, V_{\mu}^{1} \, , \, V_{\mu}^{2} \, ) \equiv(\, X_{\mu} \, , \, Y_{\mu} \,)$, and the vacuum polarizations are given by
\begin{subequations}
\begin{eqnarray}
&&
\Pi^{\mu \nu}_{11} = -i \, \eta^{\mu \nu} \, g_1^2 +
\nonumber \\
&&
\hspace{-0.8cm}
+ \mbox{Tr} \Big[ (- ig \gamma^\mu \gamma_5)S_1(p) (- ig \gamma^\nu \gamma_5) S_1(p-k) \Big]
, \;\;\;
\\
&&
\Pi^{\mu \nu}_{12} =\Pi^{\mu \nu}_{21}= -i \, \eta^{\mu \nu} \, g_3^2 \; ,
\\
&&
\Pi^{\mu \nu}_{22} = -i \, \eta^{\mu \nu} \, g_2^2 +
\nonumber \\
&&
\hspace{-0.8cm}
+ \mbox{Tr} \Big[ (- ig' \gamma^\mu \gamma_5)S_2(p) (- ig' \gamma^\nu \gamma_5) S_2(p-k) \Big] ,
\;\;\;
\end{eqnarray}
\end{subequations}
in which
\begin{subequations}
\begin{eqnarray}
S_{1}(p) = \frac{i}{\slashed{p} -m_1-\slashed{a} \, \gamma_5} \; ,
\\
S_{2}(p) = \frac{i}{\slashed{p} -m_2-\slashed{b} \, \gamma_5} \; .
\end{eqnarray}
\end{subequations}
%
%
%
%

%
Since we know that $m_{2} \gg m_{1}$ through the seesaw mechanism, we fix the $\psi_{2}$ mass at the energy scale $(\Lambda)$, {\it i.e.}, $m_{2} \sim \Lambda$. In this case, the gap equations (\ref{gapa}) and (\ref{gapb}) are simplified to $a^2 \approx 3 \pi^2 g_{1R}^2/g^2$, when
$b^\mu = 0$, and $ b^2 \approx 3 \pi^2 g_{2R}^2/g^{\prime2}$ , when $a^\mu=0$, respectively. The global minima is defined for the condition $g_{1R} > g_{2R}$ in the first gap, and $g_{2R} > g_{1R}$ in the second gap. Since experimental measurements point out to a null or very small value for $|a^2|$ which couples with the SM neutrino, we can consider the condition $g_{2R} > g_{1R}$. Under these conditions, we obtain the vacuum polarizations in the position space
%
%
\begin{widetext}
\begin{subequations}
\begin{eqnarray}
\Pi^{\mu \nu}_{11} &=& -i \, \eta^{\mu \nu} \left\{g_{1}^2 - \frac{ g^2 m_1^2}{2 \pi ^2} \left[ \frac{1}{\epsilon} - 2\ln\Big( \frac{m_{1}}{\Lambda} \Big) \right]\right\}
\nonumber \\
&&
- \frac{ig^2}{12 \pi^2}\!\Big[1- \frac{1}{\epsilon}  +2\ln\Big( \frac{m_{1}}{\Lambda} \Big) \Big]
\left(\eta^{\mu \nu} \Box-\partial^\mu \partial^\nu\right)
-\frac{ig^2}{12\pi^2} \, \partial^{\mu} \partial^{\nu} \, ,
\label{Pi11}
\\
\Pi^{\mu \nu}_{22} &=& -i \, \eta^{\mu \nu} \! \left\{g_{2}^2 - \frac{g'^2 b^2}{\pi^2}
- \frac{ g'^2 m_2^2}{2 \pi ^2} \left[ \frac{1}{\epsilon} - 2\ln\Big( \frac{m_{2}}{\Lambda} \Big) \right]
\right\}
\nonumber \\
&&
- \frac{ig^{\prime\,2}}{12 \pi^2}\left[1- \frac{1}{\epsilon}  + 2\ln\Big( \frac{m_{2}}{\Lambda} \Big)  \right](\eta^{\mu \nu} \Box-\partial^\mu \partial^\nu)
-\frac{ig'^2}{12 \pi^2} \, \partial^\mu \partial^\nu
\nonumber \\
&&
+\frac{i g'^2}{12 \pi ^2} \left(b^\mu  \partial^\nu - b^\nu \partial^\mu \right)
-\frac{ig^{\prime\,2}}{4 \pi^2} \left[ \frac{1}{\epsilon} - 2\ln\Big( \frac{m_{2}}{\Lambda} \Big) \right] \epsilon^{\mu \nu \rho\sigma} b_{\rho} \partial_{\sigma}
\nonumber \\
&&
+\frac{g'^2}{4 \pi^2} \left\{ 1 - \frac{9}{2}\left[  \frac{1}{\epsilon} - 2\ln \left(\frac{m_2}{\Lambda} \right) \right]\right\} \, b^\mu \, b^\nu \, . 
\label{Pi22}
\end{eqnarray}
\end{subequations}
\end{widetext}
Substituting these results in (\ref{Seff2}), we write the effective action as
\begin{eqnarray}
S_{eff}^{(2)}=\int d^{4}x \, {\cal L}_{eff}^{(2)} \; ,
\end{eqnarray}
where the renormalizable effective lagrangian (at the second order) is given by
\begin{eqnarray}\label{Leff2}
&&
{\cal L}_{eff}^{(2)} = -\frac{1}{4} X_{R\mu\nu}^{2}
\!-\!\frac{g_{R}^2}{24\pi^2} \, (\partial_{\mu}X_{R}^{\,\,\mu})^{2}
\!+\!\frac{1}{2} (6m_1^2) X_{R\mu}^2
\nonumber \\
&&
-\frac{1}{4} \, Y_{R\mu\nu}^2-\frac{g_{R}^{\prime\, 2}}{24\pi^2} \, (\partial_{\mu}Y_{R}^{\,\,\mu})^{2}
\!+\!\frac{1}{2} (6m_2^2) Y_{R\mu}^2
\nonumber \\
&&
+g_{3R}^2 \, X_{R\mu}Y_{R}^{\,\,\mu}
-\frac{g_{R}^{\prime \, 2}}{12\pi^2} \, (b\cdot Y_{R}) \, \partial_{\mu}Y_{R}^{\,\,\mu}
\nonumber \\
&&
-\frac{27}{4} \, (b\cdot Y_{R})^2
+\frac{3}{4} \, b_{\mu}\,\epsilon^{\mu\nu\rho\sigma}Y_{R\nu}Y_{R\rho\sigma}
\; .
\end{eqnarray}
To get the effective lagrangian in this form, we have defined the renormalized fields
\begin{eqnarray}
X_{R}^{\,\,\mu} = Z_{3}^{-1/2} X^{\mu}
\;\; , \; \;
Y_{R}^{\,\,\mu} = Z_{4}^{-1/2} Y^{\mu} \; ,
\end{eqnarray}
where the renormalization factors are given by
\begin{subequations}
\begin{eqnarray}
Z_{3}^{-1} \!\!&=&\!\! \frac{g^2}{12\pi^2}\left[1-\frac{1}{\epsilon}+ 2 \, \ln\left(\frac{m_1}{\Lambda} \right) \right] \, ,
\\
Z_{4}^{-1} \!\!&=&\!\! \frac{g^{\prime \, 2}}{12\pi^2}\left[1-\frac{1}{\epsilon}+ 2 \, \ln\left(\frac{m_2}{\Lambda} \right) \right]
\, ,
\end{eqnarray}
\end{subequations}
and the renormalized coupling constants are
\begin{eqnarray}
g_{R}\!=\!\sqrt{Z_{3}} \, g
\; , \;
g_{R}^{\prime}\!=\!\sqrt{Z_{4}} \, g^{\prime}
\; , \;
g_{3R}\!=\! \sqrt[4]{Z_{3}Z_{4}} \, g_{3} \, . \;\;
\end{eqnarray}
The transversal operators from (\ref{Pi11}) and (\ref{Pi22}) induces dynamical for the auxiliary
vector fields through the field strength tensors $X_{\mu\nu}=\partial_{\mu}X_{\nu}-\partial_{\nu}X_{\mu}$
and $Y_{\mu\nu}=\partial_{\mu}Y_{\nu}-\partial_{\nu}Y_{\mu}$, respectively.
In (\ref{Leff2}), the correspondent renormalized field strength tensors are
$X_{R\mu\nu}=Z_{3}^{-1/2}X_{\mu\nu}$, and $Y_{R\mu\nu}=Z_{4}^{-1/2}Y_{\mu\nu}$.
Furthermore, in (\ref{Leff2}), the radiative corrections also induce a new Chern-Symons
term depending on the $b^{\mu}$-parameter, $Y_{R}^{\mu}$ and the field strength tensor $Y_{R}^{\mu\nu}$.
Note that, the effective lagrangian also shows quadratic
terms in $X_{R}^{\mu}$ and $Y_{R}^{\mu}$, and also a mixed term of $X_{R}^{\mu}$
with $Y_{R}^{\mu}$. These terms are interpreted as like-massive terms of the auxiliary fields $X_{R}$ and
$Y_{R}$. We can write these terms into the matrix form
\begin{eqnarray}
{\cal L}_{mass-XY}^{(2)}=\frac{1}{2} \, \eta^{\mu\nu} (V_{R\mu})^{t} \, M^2 \, V_{R\nu} \; ,
\end{eqnarray}
in which $M^{2}$ is the mass matrix
\begin{eqnarray}
M^{2}=
\left[
\begin{array}{cc}
6m_{1}^2 & g_{3R}^2
\\
\\
g_{3R}^2 & 6m_{2}^2 \\
\end{array}
\right] \; ,
\end{eqnarray}
and $V_{R}^{\mu}$ is the column vector formed by the fields $X_{R}^{\mu}$ and $Y_{R}^{\mu}$.
This mass matrix can be diagonalized by a $SO(2)$ matrix, say ${\cal R}$, in which transformations in the vector fields are :
\begin{subequations}
\begin{eqnarray}
X_{R}^{\mu} \!&=&\! \cos\alpha \, Z_{1R}^{\mu} + \sin\alpha \, Z_{2R}^{\mu} \, ,
\label{XRrot}
\\
Y_{R}^{\mu} \!&=&\! -\sin\alpha \, Z_{1R}^{\mu} + \cos\alpha \, Z_{2R}^{\mu} \, ,
\label{YRrot}
\end{eqnarray}
\end{subequations}
where $\alpha$ is a mixing angle, such that, $\tan(2\alpha)=g_{3R}^2/(3m_{2}^2-3m_{1}^2)$. The
$Z_{1R}^{\mu}$ and $Z_{2R}^{\mu}$ are interpreted as the
physical eigenstates for the auxiliary fields whose the masses are determinate by the diagonal matrix
%
$M_{D}^{2}={\cal R}^{t} \, M^{2} \, {\cal R}=
\mbox{diag}( \, \mu_{1}^{2} \, , \, \mu_{2}^{2} \, )$, in which the eigenvalues $\left\{ \, \mu_{1}^{2} \, , \, \mu_{2}^{2} \, \right\}$ are,
respectively, read below
\begin{subequations}
\begin{eqnarray}
\mu_{1}^2 \!\!&=&\!\! 3\left(m_1^2+m_2^2\right)-\sqrt{9\left(m_1^2-m_2^2\right)^2+g_{3R}^4}
\nonumber \\
&&
\simeq 6m_{1}^2-\frac{g_{3R}^4}{6m_{2}^2} \, ,
\hspace{0.5cm}
\\
\mu_{2}^2 \!\!&=&\!\! 3\left(m_1^2+m_2^2\right)+\sqrt{9\left(m_1^2-m_2^2\right)^2+g_{3R}^4}
\nonumber \\
&&
\simeq 6m_{2}^2+\frac{g_{3R}^4}{6m_{2}^2}  \, ,
\hspace{0.5cm}
\end{eqnarray}
\end{subequations}
in which we have used the condition $m_{2}\gg m_{1}$. Since that $m_{2} \gg g_{3R}$,
we identify $\mu_{1}\simeq \sqrt{6} \, m_{1}$ as the $Z_{1R}$ mass eigenstate (light gauge boson), while that $\mu_{2}\simeq \sqrt{6} \, m_{2}$
is the correspondent eigenstate for $Z_{2R}$ (heavy gauge boson). Under this condition, we can consider
$\tan(2\alpha) \simeq g_{3R}^2/(3m_{2}^2) \ll 1$, such that $\sin\alpha\simeq g_{3R}^2/(6m_{2}^2) \simeq 0$ and $\cos\alpha \simeq 1$. In the $\left\{ \, Z_{1R} \, , \, Z_{2R} \, \right\}$ basis, the effective lagrangian (\ref{Leff2}) is
%
%
%
\begin{eqnarray}\label{Leff2Z1Z2}
&&
{\cal L}_{eff}^{(2)} = -\frac{1}{4} Z_{1R\mu\nu}^{2}
+\frac{1}{2} \, \mu_{1}^2 \, Z_{1R\mu}^2
\!-\!\frac{g_{R}^2}{24\pi^2} \, (\partial_{\mu}Z_{1R}^{\,\,\mu})^{2}
\nonumber \\
&&
-\frac{1}{4} \, Z_{2R\mu\nu}^2+\frac{1}{2} \, \mu_{2}^{2} \, Z_{2R\mu}^2
-\frac{g_{R}^{\prime\, 2}}{24\pi^2} \, (\partial_{\mu}Z_{2R}^{\,\,\mu})^{2}
\nonumber \\
&&
-\frac{g_{R}^{\prime\,2}}{12\pi^2}
\,(b\cdot Z_{2R})\,\partial_{\mu}Z_{2R}^{\,\,\mu}
-\frac{27}{4} \,
(b\cdot Z_{2R})^2
\nonumber \\
&&
+\frac{3}{4} \, b_{\mu}\,\epsilon^{\mu\nu\rho\sigma}
\,Z_{2R\nu} \, Z_{2R\rho\sigma}
\, .
\end{eqnarray}

%
Notice that the Chern-Symons term emerges for the $Z_{2R}^{\,\,\mu}$ physical eigenstate when the mixing with the $Z_{1R}^{\,\,\mu}$ is very weak.
The action principle yields the field equations :
\begin{subequations}
\begin{eqnarray}
&&
\partial_{\mu}Z_{1R}^{\mu\nu} + \mu_{1}^2 \, Z_{1R}^{\nu}
\!+\!\frac{g_{R}^2}{12\pi^2} \, \partial^{\nu}(\partial_{\mu}Z_{1R}^{\,\,\mu})=0
\, , \hspace{0.5cm}
\label{eqZ1R}
\\
&&
\partial_{\mu}Z_{2R}^{\mu\nu} + \mu_{2}^2 \, Z_{2R}^{\nu}
\!+\!\frac{g_{R}^{\prime\,2}}{12\pi^2} \, \partial^{\nu}(\partial_{\mu}Z_{2R}^{\,\,\mu})
\nonumber \\
&&
-\frac{g_{R}^{\prime\,2}}{12\pi^2}\left[ \, b^{\nu} \, \partial_{\mu}Z_{2R}^{\,\,\mu}- \partial^{\nu}(b\cdot Z_{2R}) \right]
\nonumber \\
&&
+\frac{3}{2} \, \epsilon^{\mu\nu\rho\sigma} \, b_{\mu} \, Z_{2R\rho\sigma}=0 \, .
\label{eqZ2R}
\end{eqnarray}
\end{subequations}
Operating $\partial_{\nu}$ on the equations (\ref{eqZ1R}) and (\ref{eqZ2R}), the longitudinal part of
$Z_{1R}$ and $Z_{2R}$ satisfies the relations
\begin{subequations}
\begin{eqnarray}
&&
\left(\Box+\frac{12\pi^2\mu_{1}^2}{g_{R}^2} \right)\partial_{\mu}Z_{1R}^{\mu}=0 \; ,
\\
&&
\left(\Box-b\cdot\partial+\frac{12\pi^2\mu_{2}^2}{g_{R}^{2\,\prime}} \right)\partial_{\mu}Z_{2R}^{\mu}+\Box(b\cdot Z_{2R})=0 \; .
\nonumber \\
\end{eqnarray}
\end{subequations}
Substituting the plane wave solutions $Z_{iR}^{\mu}(x)=z_{iR}^{\mu} \, e^{ik\cdot x} \, (i=1,2)$,
in which $z_{iR}^{\mu}$ are the constant and uniform wave amplitudes, the equations (\ref{eqZ1R}) and (\ref{eqZ2R})
can be written as $M_{1}^{\mu\nu}\,z_{1R\mu}=0$ and $M_{2}^{\mu\nu}\,z_{2R\mu}=0$, respectively,
where the matrices $M_{i}^{\mu\nu} \, (i=1,2)$ are given by
\begin{subequations}
\begin{eqnarray}
M_{1}^{\mu\nu} \!\!&=&\!\! \left(-k^2+\mu_{1}^2\right) \eta^{\mu\nu}+\left(1- \frac{g_{R}^2}{12\pi^2} \right) k^{\mu}k^{\nu} \; ,
\hspace{0.8cm}
\\
M_{2}^{\mu\nu} \!\!&=&\!\! \left(-k^2+\mu_{2}^2\right) \eta^{\mu\nu}+\left(1- \frac{g_{R}^{\prime\,2}}{12\pi^2} \right) k^{\mu}k^{\nu}
\nonumber \\
&&
\hspace{-0.5cm}
+\frac{ig_{R}^{\prime\,2}}{12\pi^2}\left(b^{\mu}k^{\nu}-b^{\nu}k^{\mu}\right)+i \, 3 \, \epsilon^{\mu\nu\rho\sigma} \, b_{\rho} \, k_{\sigma} \; .
\end{eqnarray}
\end{subequations}
The dispersion relations are determinate by the conditions $\mbox{det}(M_{1}^{\mu\nu})=0$ and
$\mbox{det}(M_{2}^{\mu\nu})=0$. The null determinant of $M_{1}^{\mu\nu}$ yields the frequencies solutions
$\omega_{Z_{1R}}^{(i)\pm}=\pm \, \omega_{Z_{1R}}^{(i)}({\bf k}) \, (i=1,2)$, in which the dispersion relations
$\omega_{Z_{1R}}^{(i)}({\bf k})$ are read below :
\begin{subequations}
\begin{eqnarray}
\omega_{Z_{1R}}^{(1)}({\bf k}) \!&=&\! \sqrt{{\bf k}^2+\mu_{1}^2} \; ,
\label{omega1Z1R}
\\
\omega_{Z_{1R}}^{(2)}({\bf k}) \!&=&\! \sqrt{{\bf k}^2+\frac{12\pi^2\mu_{1}^2}{g_{R}^2}} \; .
\label{omega2Z1R}
\end{eqnarray}
\end{subequations}
The frequency (\ref{omega1Z1R}) is the transversal mode of $Z_{1R}$ with mass $\mu_{1}=\sqrt{6} \, m_{1}$,
and (\ref{omega2Z1R}) is the frequency associated with longitudinal propagation mode of $Z_{1R}$.
For the case of $M_{2}^{\mu\nu}$, we assume a time-like vector $b^{\mu}=(b_{0},{\bf 0})$ in which
four roots are possible for the null determinant :
$\omega_{Z_{2R}}^{(i)\pm}=\pm \, \omega_{Z_{2R}}^{(i)}({\bf k}) \, (i=1,2,3,4)$.
These frequencies are given by
\begin{widetext}
\begin{subequations}
\begin{eqnarray}
\omega_{Z_{2R}}^{(1)}({\bf k}) \!\!&=&\!\! \sqrt{{\bf k}^2-3|b_{0}||{\bf k}|+\mu_{2}^2} \; ,
\\
\omega_{Z_{2R}}^{(2)}({\bf k}) \!\!&=&\!\! \sqrt{{\bf k}^2+3|b_{0}||{\bf k}|+\mu_{2}^2} \; ,
\\
\omega_{Z_{2R}}^{(3)}({\bf k}) \!\!&=&\!\! \sqrt{ {\bf k}^2+\frac{ 3\pi(g_{R}^{\prime\,2}+12\pi^2)\mu_{2}^2-\sqrt{ 9\pi^2(g_{R}^{\prime\,2}-12\pi^2)^2\mu_{2}^{4}-3b_{0}^2 \, g_{R}^{\prime \, 6} \, {\bf k}^2 } }{ 6\pi g_{R}^{\prime\,2} }  } \; ,
\\
\omega_{Z_{2R}}^{(4)}({\bf k}) \!\!&=&\!\! \sqrt{ {\bf k}^2+\frac{ 3\pi(g_{R}^{\prime\,2}+12\pi^2)\mu_{2}^2+\sqrt{ 9\pi^2(g_{R}^{\prime\,2}-12\pi^2)^2\mu_{2}^{4}-3b_{0}^2 \, g_{R}^{\prime \, 6} \, {\bf k}^2 } }{ 6\pi g_{R}^{\prime\,2} }  } \; .
\end{eqnarray}
\end{subequations}
\end{widetext}
Notice that, when $b_{0} \rightarrow 0$, the previous frequencies reduce to (\ref{omega1Z1R}) and $(\ref{omega2Z1R})$
exchanging $g_{R}^{\prime} \rightarrow g_{R}$ and $\mu_{2} \rightarrow \mu_{1}$.
Finally, we obtain the $n=3$ contribution to the effective action, under the condition $a^{\mu}=0$ and $b^{\mu} \neq 0$, is :
\begin{subequations}
\begin{eqnarray}
\Pi^{X}_{\;\;\mu \nu \rho} \!\!&=&\!\! 0 \; ,
\\
\Pi^{Y}_{\;\;\mu \nu \rho} \!\!&=&\!\! \frac{i g^{\prime \, 3} }{3 \pi^2} \left( \beta_\mu \, \eta_{ \nu \rho} + \beta_\nu \, \eta_{\nu \rho} + \beta_\rho \, \eta_{\mu \nu} \right) . \;\;
\end{eqnarray}
\end{subequations}
%
%
The divergent contribution vanishes due to the index symmetry.  For the $n=4$ contribution, one finds:
\begin{subequations}
\begin{eqnarray}
\Pi^{X}_{\;\;\mu \nu \rho \kappa} \!\!&=&\!\! \frac{i g^4 }{3 \pi^2} \left( \eta_{\mu \nu} \eta_{\rho \kappa} - \eta_{\mu \rho} \eta_{\nu \kappa} + \eta_{ \mu \kappa} \eta_{\nu \rho} \right) , \hspace{0.8cm}
\\
\Pi^{Y}_{\;\;\mu \nu \rho \kappa} \!\!&=&\!\! \frac{i g^{\prime \, 4} }{3 \pi^2} \left( \eta_{\mu \nu} \eta_{\rho \kappa} - \eta_{\mu \rho} \eta_{\nu \kappa} + \eta_{ \mu \kappa} \eta_{\nu \rho} \right) . \hspace{0.8cm}
\end{eqnarray}
\end{subequations}
%
Again, the divergence vanishes due to the index symmetry. Therefore, using these results, the renormalized effective potential
can be written in terms of the physical eigenstates $\{ \, Z_{1R} \, , \, Z_{2R} \, \}$ as :
\begin{eqnarray}
&&
V_{eff}(Z_{1R},Z_{2R})= - \frac{1}{2} \, \mu_1^2 \, Z_{1R}^2  + \frac{g_R^4}{12 \pi^2} \left( Z_{1R}^2 \right)^2
\nonumber \\
&&
- \frac{1}{2} \, \mu_2^2 \, Z_{2R}^2
+\frac{(g'_R)^4}{12 \pi^2} \left( Z_{2R}^2 +\frac{2}{g'_R}Z_{2R} \cdot b \right)^2 \!\!. \,\,
\end{eqnarray}
where we have fixed the condition $a^{\mu} = 0$, with the shift $B^\mu \rightarrow \frac{b^\mu}{g'_R} + Y^\mu$, and
the rotation to the physical eigenstates from (\ref{XRrot}) and (\ref{YRrot}).
%

%
\section{Neutrino Oscillations in the DLSB scenario}
\label{sec5}
Since the discovery of the oscillation phenomena associated with the non-null neutrino's masses in a basis which does not match with the flavor basis of the SM, several models beyond the SM were proposed in order to explain this phenomena in the literature \cite{KanemuraPRD2014,ColomaJHEP2021,BonillaPRD2020}. We start the description of the oscillation phenomena considering the three flavor neutrinos eigenstates $| \nu_{\alpha} \rangle =\left\{ \, |\nu_e\rangle \, , \, |\nu_\mu\rangle \, , \, |\mu_\tau\rangle \, \right\}$. After the seesaw mechanism takes place, we define the physical eigenstates $| \nu_i \rangle$, with $i=\left\{ \, 1, \, 2 \, , \, 3 \, \right\}$, that represent the LHNs with light masses. These physical eigenstates are related with the previous flavour neutrinos eigenstates by the transformation

\begin{equation}
| \nu_i \rangle  = \sum_\alpha U_{i \alpha} \, | \nu_\alpha \rangle ~~~,~~~ | \nu_\alpha \rangle  = \sum_i U^{\ast}_{ \alpha i } \, | \nu_i \rangle \; ,
\end{equation}
in which $U_{i \alpha}$ is the so-called Pontecorvo–Maki–Nakagawa–Sakata (PMNS) unitary and complex matrix. The dynamic of the physical eigenstates is ruled by the spatial-time evolution:
\begin{equation}
|\nu_i({\bf x}, t) \rangle = e^{i(E_i t + {\bf p}_i \cdot {\bf x})} \, |\nu_i(0,0) \rangle \; .
\end{equation}
Therefore, the transition probability of oscillation between flavors states $| \nu_\alpha \rangle \rightarrow | \nu_\beta \rangle$ is defined by
\begin{eqnarray}
P_{\alpha \rightarrow \beta}({\bf x}, t) = |\langle \nu_\alpha ({\bf x},t)| \nu_\beta({\bf x},t) \rangle |^2
\nonumber \\
=\Bigg{|} \sum_{i} U^{\ast}_{i \alpha} \, U_{i \beta} \, e^{i E_i t} \, e^{i {\bf p}_i \cdot {\bf x}}\Bigg{|}^2 \; .
\end{eqnarray}
The energy of a $i$-neutrino in the physical eigenstate basis is
\begin{eqnarray}
E_i = \sqrt{{ \bf p}_i^2 + m_i^2} \simeq |{\bf p}_i| + \frac{m_i^2}{2|{\bf p}_i|} = E + \frac{m_i^2}{2 E} \; , \;\;
\end{eqnarray}
where we have considered the approximation $|{\bf p}_{i}| \gg m_{i}$, and assuming that $t \approx x$ and ${\bf p}_i -{\bf p}_j \approx 0$,
the transition probability is :
\begin{widetext}
\begin{eqnarray}\label{Palphabeta}
P_{\alpha \rightarrow \beta}(x)
\!\!&=&\!\! \delta_{\alpha\beta}-4\sum_{\substack{i,j=1,\\ i > j}}^{3} \,
\Re[U^{\ast}_{i \alpha} \, U_{i \beta} \, U_{\alpha j} \, U^{\ast}_{\beta j}]
\sin^2\left[\left(\frac{\Delta m_{ij}^{\,2}}{4 E}\right) x\right]+
\nonumber \\
&&
\hspace{-0.5cm}
+\,2\sum_{\substack{i,j=1,\\ i > j}}^{3} \,
\Im[U^{\ast}_{i \alpha} \, U_{i \beta} \, U_{\alpha j} \, U^{\ast}_{\beta j}]
\sin\left[\left(\frac{\Delta m_{ij}^{\,2}}{2 E}\right) x\right]
\; ,
\end{eqnarray}
\end{widetext}
in which $\Re$ and $\Im$ are the real and imaginary parts of the PMNS matrix elements.
The previous result (\ref{Palphabeta}) is associated with the usual dispersion relation for the light $i$-neutrino.
Notice also that this result depends on the PMNS matrix elements, and depends on the squared difference of the mass
between a $i$-physical eigenstate and a $j$-physical eigenstate, $\Delta m_{ij}^{2}:=m_{i}^2-m_{j}^2$, and the known oscillation
length is $\ell_{ij}=4\pi E/|m_{i}^2-m_{j}^2|$.
In the case of the model with DLSB, we consider the transition probability for the neutrino dispersion relation from the phase I in (\ref{omega1m1peq}),
in which the VEV parameter is time-like, {\it i.e.}, $a^{\mu}=(a^{0},{\bf 0})$. In this particular case, the light $i$-neutrino
dispersion relation is :
\begin{eqnarray}
E_i^{(\pm)} \!\!&=&\!\!  \sqrt{||{ \bf p}_i| \pm \mu_i|^2 + m_i^2}
\nonumber \\
&\simeq& ||{\bf p}_i|\pm \mu_i| + \frac{m_i^2}{2 ||{\bf p}_i|+\mu_i|}
\nonumber \\
&\simeq& |E \pm \mu_i | + \frac{m_i^2}{2 |E\pm  \mu_i|} \; ,
\end{eqnarray}
where $\mu_i := (a^{0})_{i}=\pi\sqrt{3}\,(g_{1R})_{i}/g \, (i=1,2,3)$ means the time-like parameter for each $i$-neutrino (or $i$-antineutrino), $E_{i}^{(+)}$ is the energy of the light $i$-neutrino, and $E_{i}^{(-)}$ is the correspondent for the $i$-antineutrino. To obtain the difference of energy of a $i$-neutrino with a $j$-neutrino in this scenario, we assume $\left( \, \mu_{i} \, , \, \mu_{j} \, \right)>0$, and with a flavour dependent on the coupling constant $(g_{1R})_{i} \, (i=1,2,3)$. Under these conditions, the difference of energy is
\begin{eqnarray}
E_i^{(\pm)} - E_j^{(\pm)} &\simeq& |E \pm \mu_i | - |E \pm \mu_j |
+
\nonumber \\
&& \hspace{-0.5cm}
+ \frac{m_i^2}{2 |E\pm  \mu_i|}- \frac{m_j^2}{2 |E\pm  \mu_j|}
\nonumber \\
&&\hspace{-0.5cm}\approx \pm \left(\mu_i - \mu_j\right) + \frac{\Delta m_{ij}^2}{2 E} \; ,
\end{eqnarray}
where we have used $E \gg \left( \, \mu_i \, , \, \mu_j \, \right)$ in the last line.
Notice that this result yields different oscillation lengths for the neutrino and antineutrino
induced by the DLSB violation.
%

\begin{figure}[H]
\centering
\includegraphics[scale=0.51]{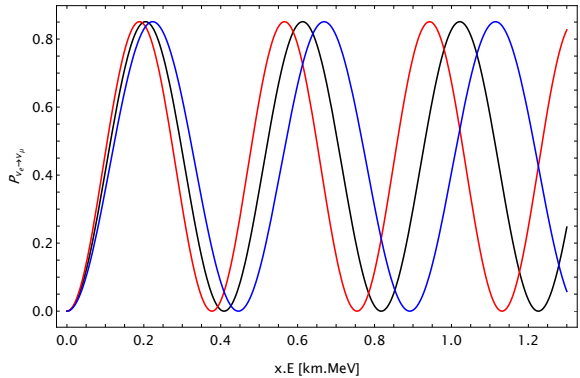}
\caption{The transition probability $P_{\nu_e \rightarrow \nu_\mu}$ as function of the distance $x$ in energy $(E)$ units.
We use $\Delta m_{12}^2 = 0.041 \, \mbox{eV}^2$
and $\sin^2(2\theta_{12}) = 0.96$, for the values $(\Delta g_{1R})_{12}/g = 0$ (black line), $(\Delta g_{1R})_{12}/g = 0.1 \,\mbox{meV}$ (red line) and $(\Delta g_{1R})_{12}/g = -0.1 \, \mbox{meV}$ (blue line).}
\label{osc1}
\end{figure}

\begin{figure}[H]
\centering
\includegraphics[scale=0.51]{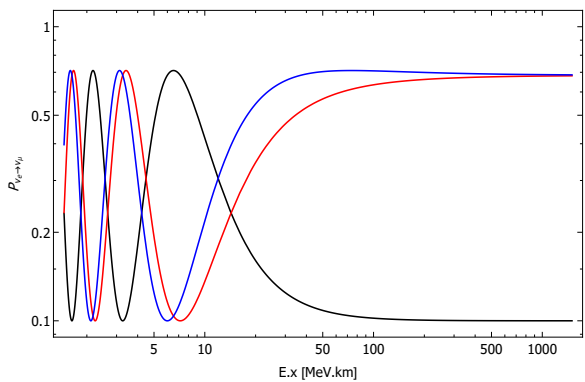}
\caption{The transition probability $P_{\nu_e \rightarrow \nu_\mu}$ as function of the energy $(E)$. We also choose
$\Delta m_{12}^2 = 0.041 \, \mbox{eV}^2$ and $\sin^2(2\theta_{12}) = 0.96$,
for the values $(\Delta g_{1R})_{12}/g = 0$ (black line), $(\Delta g_{1R})_{12}/g =  0.2 \, \mbox{eV}$ (red line),
and $(\Delta g_{1R})_{12}/g = - 0.2 \, \mbox{eV}$ (blue line).}
\label{osc2}
\end{figure}

%
To simplify our future estimative, we reduce our problem for the case of the transition probability of muon and electron neutrinos.
Thereby, the PMNS matrix is reduced in terms of the electron-muon mixing angle $\theta_{12}$ :
\begin{equation}
\left(
\begin{array}{c}
\nu_{1} \\
\\
\nu_{2} \\
\end{array}
\right)
=
\left[
\begin{array}{cc}
\cos\theta_{12}  & \sin\theta_{12} \\
\\
-\sin\theta_{12} & \cos\theta_{12} \\
\end{array}
\right]
\left(
\begin{array}{c}
\nu_{e} \\
\\
\nu_{\mu} \\
\end{array}
\right)
 \; ,
\end{equation}
where $\sin^2\theta_{12}=0.307 \pm 0.013$ \cite{PDG2020}. The probability of the $\nu_{e}$-neutrino changes its flavour to $\nu_{\mu}$ is read below :
\begin{eqnarray}
P_{\nu_e \rightarrow \nu_\mu}(x) \!\!&=&\!\! \sin^2 (2\theta_{12}) \sin^2 \left[\left(E_{1}^{(+)}-E_{2}^{(+)}\right)\frac{x}{2} \right]
\nonumber \\
\!\!&=&\!\!\sin^2 (2\theta_{12}) \sin^2\left(\frac{\pi x}{\ell}\right)
\; ,
\end{eqnarray}
in which the correspondent oscillation length $\ell$ is
\begin{eqnarray}
\ell^{-1}&=&\left|\frac{\pi \, \sqrt{3}}{2g} \, (\Delta g_{1R})_{12}+\frac{\Delta m_{12}^2}{4\pi E}\right| \; ,
\hspace{0.3cm}
\end{eqnarray}
where $(\Delta g_{1R})_{12}:=(g_{1R})_{1}-(g_{1R})_{2}$,
and the difference of squared masses is in the range $\Delta m_{12}^{2} = 10^{-2} - 10^{-3} \, \mbox{eV}^{2}$ for experiments
in Accelerators \cite{PDG2020}. The usual oscillation length in the literature is recovered when $(g_{1})_{i} \rightarrow 0$.
For a strong coupling regime (SCR) $(\Delta g_{1R})_{12} \gg \Delta m_{12}^2/E$, we obtain
\begin{equation}\label{lengthapp}
\ell^{-1} \simeq \pi \, \frac{\sqrt{3}}{2} \, \left|\frac{(\Delta g_{1R})_{12}}{g}\right| \; .
\end{equation}
The transition probability $P_{\nu_{e}\rightarrow\nu_{\mu}}$ is illustrated as a function of $x$ in fig. \eqref{osc1}, and posteriorly, it is also illustrated as function of energy $(E)$ in fig. \eqref{osc2}, respectively. In both plots, we have used the $\Delta m_{12}^2 = 0.041 \, \mbox{eV}^2$ and $\sin^2(2\theta_{12}) = 0.96$ \cite{Arevalo2}. In the fig. (\ref{osc1}), we have chosen the values $(\Delta g_{1R})_{12}/g = 0$ (black line), $(\Delta g_{1R})_{12}/g = +0.1 \, \mbox{meV}$ (red line) and $\Delta G = -0.1 \, \mbox{meV}$ (blue line), in energy $(E)$ units . In fig. (\ref{osc2}), we have chosen $(\Delta g_{1R})_{12}/g = 0$ (black line), $(\Delta g_{1R})_{12}/g =  0.2 \, \mbox{eV}$ (red line), and $(\Delta g_{1R})_{12}/g = - 0.2 \, \mbox{eV}$ (blue line). The graphics show that the effect of the DLSB in the neutrino oscillation is greater for large energy values.


%
%
\begin{figure}[H]
\centering
\includegraphics[scale=0.51]{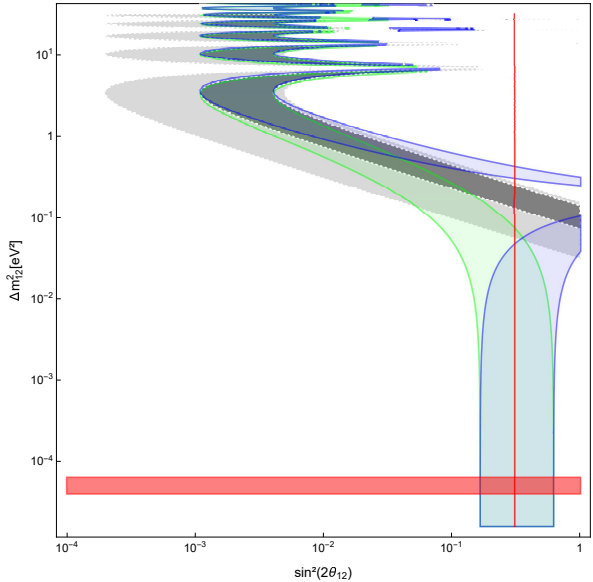}
\caption{The parameter space of $\mbox{sin}^2(2 \theta_{12})$ versus $\Delta m_{12}^2$ for the eletron-neutrino oscillation. The gray region shows the region with $68\%$ confidence level and the light gray shows the region for the $95\%$ confidence level, according to the LSND experiment. The green and blue regions shows the allowed region for $(\Delta g_{1R})_{12}/g = 10^{-9} \, \mbox{eV}$ and $(\Delta g_{1R})_{12}/g = - 10^{-9} \, \mbox{eV}$ based on the LSND result ($68\%$ c.l.). The red region indicates the region with $68\%$ confidence level from the Super-Kamiokande experiment \cite{PDG2020}.}
\label{paramplot}
\end{figure}

As can be seen in figure \eqref{osc1}, the DLSB effect is to shift the transition probability in the $x$-axis. Notice that, in fig. \eqref{osc2}, the effect of the DLSB in the neutrino oscillation is to generate oscillation even for high energies, in opposition to the usual result (black line). The red (neutrinos) and blue (anti-neutrinos) lines also show oscillation, but the curves go to a finite value of probability that is contributed by the DLSB parameter. For large energy, the oscillation length is given by (\ref{lengthapp}). If we consider $(\Delta g_{1R})_{12}/g =  \pm \, 0.2 \, \mbox{eV}$,
the transition probability limit $(E\cdot x \gg 1)$ in the red and blue curves from the fig. (\ref{osc2}) is
\begin{eqnarray}
P_{\nu_{e}\rightarrow\nu_{\mu}} (E \cdot x \gg 1) \approx 0.65 \; .
\end{eqnarray}
%
%
Going further, in fig. \eqref{paramplot}, we show the allowed region in the parameter space of $\mbox{sin}^2(2 \theta_{12})$ versus $\Delta m_{12}^2$.
Based on an oscillation probability of $P_{\nu_e \rightarrow \nu_\mu} = (2.6 \pm 1.5) \times 10^{-3}$ from the LSND experiment \cite{AthanassopoulosPRL96}, and using $E = 60 - 200 \, \mbox{MeV}$ and $x=L \approx 30 \, \mbox{m} = 2.4 \times 10^7 \, \mbox{eV}^{-1}$, we plot the range of $\sin^2(2 \theta_{12})$ and $\Delta m_{12}^2$ values compatible with this result (gray and light gray regions in fig. \eqref{paramplot}, for 68 $\%$ C.L. and 95 $\%$ C.L., respectively).
Based on the phenomenological bounds from \cite{kost1}, we consider $(\Delta g_{1R})_{12}/g \approx \pm \, 10^{-10} \, \mbox{eV}$, and plotting their contribution in the probability one can see in the figure \eqref{paramplot} where the green and blue regions are in contact with long base line (LBL) accelerators results (interception of the red regions), in contrast to the symmetric case (gray regions).
It happens due to the smallness of the argument $\frac{\Delta m_{12}^2}{4\pi E}$, since that $\frac{\Delta m_{12}^2}{4\pi E} \approx 10^{-12} \mbox{eV}$. Thereby, the contribution of the mass splitting and DLSB parameter is small and obeys the hierarchy condition $\frac{\pi \, \sqrt{3}}{2g} \, (\Delta g_{1R})_{12}>\frac{\Delta m_{12}^2}{4\pi E}$, such way the oscillation length in the eq. \eqref{lengthapp} can be rewritten as $\frac{1}{\ell} \approx \frac{\sqrt{3} \pi}{2g} \,  \, \left(\Delta g_{1R}\right)_{12}$. Finally, one can note that $L/\ell < 1$, and due to this property, one can approximate the oscillation probability as:
%
\begin{equation}
P^{SCR}_{\nu_e \rightarrow \nu_\mu}  \approx   \frac{3 \pi^4}{8}\sin^2(2 \theta_{12}) \left[ \,  L \, \frac{(\Delta g_{1R})_{12}}{g} \, \right]^2\; ,
\end{equation}
%
and using the PDG best fit for $\mbox{sin}^2(2 \theta_{12}) \approx 0.31$ and $L \approx 30 \, \mbox{m}$, one finds:
\begin{equation}
P^{SCR}_{\nu_e \rightarrow \nu_\mu}  \approx  6.5 \times 10^{15} \, \Bigg[ \, \frac{(\Delta g_{1R})_{12}}{g} \, \Bigg]^2 \!\! = (2.6 \, \pm \, 1.5) \times 10^{-3} \; .
\end{equation}
%
Therefore, one can infer that $\Big|\frac{(\Delta g_{1R})_{12}}{g}\Big| = \, 8 \times 10^{-19}$ GeV with 68$\%$ C.L., the same order of magnitude of the bounds from the tandem model \cite{Katori}, and consistent with our initial assumptions.
%
%

%


%
\section{Conclusions}
\label{sec6}
A model with two massive Majorana fermions coupled through self-quartic interactions themselves was proposed in this paper. After a type-II seesaw mechanism, one of the Majorana fermions acquires a light mass $(m_1)$, the other one gains a heavy mass $(m_{2})$, and we obtain an effective model with quartic self-interaction fermion theory of axial currents in the physical basis. The model allows the introduction of two auxiliary gauge fields in which the effective potential has a minimum at two vacuum expected values (VEVs) that are constant $4$-vectors. These VEVs scales break dynamically the Lorentz symmetry of the model, and as consequence, the dispersion relations of the Majorana fermions are modified.
We calculate the correspondent frequencies solutions for the light and heavy Majorana fermions in the scenarios of time-like and space-like $4$-vectors.
Posteriorly, we analyze the fluctuations of the auxiliary gauge fields around the VEV scales to calculate the effective action expanded up to second order. Therefore, the radiative corrections yield kinetic terms and mixed mass terms for the gauge fields. The mass matrix is so diagonalized, where the physical eigenstates have one mass of $M_{Z_{1R}} \approx \sqrt{6} \, m_{1}$ for the light gauge field $(Z_{1R})$, and a mass of $M_{Z_{2R}} \approx \sqrt{6} \, m_{2}$ for the heavy gauge field $(Z_{2R})$. Since we just consider one DLSB parameter, the radiative corrections also generate the Chern-Symons term associated with the heavy gauge field. Using the field equations, we obtain the dispersion relations and the frequency solutions for the gauge fields on a diagonal basis.
To end, we investigate the DLSB in the neutrino's oscillation. The dispersion relation obtained previously for the light Majorana fermions is used to calculate the transition probability $\nu_{e} \rightarrow \nu_{\mu}$ and the length oscillation. It is well-known that the neutrino oscillation in the electron-muon sector measured by Super-Kamiokande experiment \cite{Fukuda} differs from the LSND \cite{AthanassopoulosPRL96,AthanassopoulosPRL98} and the MiniBooNe experiments \cite{Arevalo2}, as can be seen in fig. \eqref{phasespace}, where one plots the parameter space of $\mbox{sin}^2(2\theta_{12})$ versus $\Delta m_{12}^{2}$. The allowed region for the LSND and the Super-Kamiokande experiments can be seen as the red and gray regions of \eqref{phasespace}, respectively. The blue and green regions represents the allowed phase space with the contribution of the DLSB parameter with approximate value of $\frac{(\Delta g_{1R})_{12}}{g} \approx \pm \, 6 \times 10^{-19}$ GeV on the DLSB parameter.
As can be seen in fig. \eqref{phasespace}, if we admit the light and heavy sector interacts through the axial channel ($G_3 \neq 0$) only one
of the sectors violates dynamically the Lorentz symmetry, and the VEVs scales are given by
\begin{eqnarray}
a^2 = \frac{3 \pi^2 \, G_2}{G_1 G_2-G_3^2} \hspace{0.5cm} \mbox{and} \hspace{0.5cm} b^2=0 \; ,
\end{eqnarray}
or
\begin{eqnarray}
a^2=0
\hspace{0.5cm} \mbox{and} \hspace{0.5cm}
b^2 = \frac{3 \pi^2 \, G_1}{G_1 G_2-G_3^2} \; ,
\end{eqnarray}
where $a^{\mu}$ and $b^{\mu}$ couple with the light and heavy sectors, respectively. Thereby, there is a possibility in the phase diagram for the DSLB occurs in the heavy sector maintaining the standard model of the symmetric sector under Lorentz symmetry. One also has that for $G_3 = 0$, {\it i.e.}, in the decoupled scenario, both the VEVs scales can be non-null simultaneously.
If one assumes $G_3 = 0$, {\it i.e.}, the light neutrino sector decouples from the heavy sector, one can estimate the Lorentz violation parameter in the axial neutrino self-interaction. The authors choose this framework since the heavy sector up to date was not detected, which indicates that, if it exists, the heavy sector interacts weakly with the SM particles, particularly with neutrinos. 
Using the SME limits from  $a^0_e = |(a_L)^T_{ee}| < 2 \times 10^{-27} \, \mbox{GeV}$ (from table D29 of ref. \cite{SME2021}), we obtain the upper bound :
\begin{equation}
|(a_L)_{\mu \mu}^T| \lesssim 3 \times 10^{-18} \, \mbox{GeV} \;  ,
\end{equation}
which means the muonic neutrino sector would be the relevant sector in the axial quartic interactions. Based on the experimental results from oscillations of muonic neutrinos \cite{minos,icecube}, which reveals no discrepancies with the standard neutrino oscillation models, one can roughly assume the following hierarchy:
\begin{equation}
|(a_L)_{ee}^T| \ll |(a_L)_{\mu \mu}^T| \approx |(a_L)_{\tau \tau}^T| \; .
\end{equation}
Is important to comment on the case $G_3 \neq 0$: If one assumes a more general statement that the light and heavy sectors interact through $G_3$, the analysis become much more complex. In fact, any attempt to find any bound assuming $G_3 \neq 0$ becomes unfeasible with the actual experimental data. In addition to the fact that $G_2$ and $G_3$ are constants related to the heavy sector (which one has no information about) more general assumptions should be made, {\it i.e.}, the full flavor structure of the quartic interactions. These features will be studied in other opportunities.

From the phenomenological point of view, in the high energy limit, the quantity $\Delta m^2/E$ vanishes and implies that neutrinos
at high energy do not oscillate. In the case of DLSB, the oscillation remains even in the high energy limit and can be a motivation
to search for oscillation patterns in events from natural astrophysical phenomena which produce particles with energy beyond the PeV scale
\cite{icecubePeV,IceCubeNature}.
Going further, since $|a^0| \propto G_1^{-1/2}$, one can roughly estimate the coupling constant for the muon and tau neutrinos : $G_{1}^{\,\,(\mu)}
\approx G_{1}^{\,\,(\tau)} \approx 10^{34} \, \mbox{GeV}^{-2}$. In such a strong coupling environment non-perturbative phenomena may take place and non-linear aspects could drive the system \cite{YuriPRD2021}. The effects of this kind of interaction also could be tested in the context of supernova processes and could also generate cosmological implications. These new features will be research subjects in forthcoming papers.

\section*{Acknowledgments}
Y.M.P.G. is supported by a postdoctoral grant from  {}Funda\c{c}\~ao
Carlos Chagas Filho de Amparo \`a Pesquisa do Estado do Rio de Janeiro
(FAPERJ). M. J. Neves thanks CNPq (Conselho Nacional de Desenvolvimento Cient\' ifico e Tecnol\'ogico), Brazilian scientific support federal agency, for partial financial support, Grant number 313467/2018-8.
%


%

%

\begin{thebibliography}{99}
%
%
\bibitem{Araki} T. Araki et al. (KamLAND Collaboration), {\it Measurement of Neutrino Oscillation with KamLAND: Evidence of Spectral Distortion},
Phys. Rev. Lett. {\bf 94}, 081801 (2005).

%
\bibitem{AthanassopoulosPRL96} C. Athanassopoulos et al, Phys. Rev. Lett. {\bf 77}, 3082 (1996).
%
\bibitem{AthanassopoulosPRL98} C. Athanassopoulos et al, Phys. Rev. Lett. {\bf 81}, 1774 (1998).
%

\bibitem{Fukuda} Y. Fukuda et al. (Super-Kamiokande Collaboration), {\it Evidence for Oscillation of Atmospheric Neutrinos}, Phys. Rev. Lett. {\bf 81}, 1562 (1998).

\bibitem{Mention} G. Mention, M. Fechner, Th. Lasserre, Th. A. Mueller, D. Lhuillier, M. Cribier and A. Letourneau, {\it The Reactor Antineutrino Anomaly}, Phys. Rev. D {\bf 83}, 073006 (2011).

\bibitem{Aguilar} A. Aguilar et al, {\it Evidence for Neutrino Oscillations from the Observation of Electron Anti-neutrinos in a Muon Anti-Neutrino Beam}, Phys. Rev. D {\bf 64}, 112007 (2001).

\bibitem{Smirnov} M.V. Smirnov, Zh.J. Hu, J.J. Ling, Yu.N. Novikov, Z. Wang and G.
Yang, {\it Sterile neutrino oscillometry with Jinping}, The European Physical Journal C {\bf 80}, 609 (2020).

\bibitem{Arevalo} A. A. Aguilar-Arevalo et al (The MiniBooNE Collaboration), {\it A Combined $\nu_{\mu} \rightarrow \nu_{e}$ \& $\bar{\nu}_{\mu} \rightarrow \bar{\nu}_{e}$ Oscillation Analysis of the MiniBooNE Excesses},  arXiv:hep-exp/1207.4809v2.


\bibitem{Arevalo2} A. A. Aguilar-Arevalo et al (The MiniBooNE Collaboration), {\it Significant Excess of Electronlike Events in the MiniBooNE Short-Baseline Neutrino Experiment}, Phys. Rev. Lett. {\bf 121}, 221801 (2018).


%
%
%

\bibitem{seesaw1} P. Minkowski, {\it $\mu \rightarrow e \, \gamma$ at a rate of one out of $10^{9}$ muon decays?}, Phys. Lett. B {\bf 67}, 421 (1977).

\bibitem{seesaw2} R. N. Mohapatra and G. Senjanovi\'{c}, {\it Neutrino Mass and Spontaneous Parity Nonconservation},
Phys. Rev. Lett. {\bf 44}, 912 (1980).

 \bibitem{SenjanovicPRD} R. N. Mohapatra and G. Senjanovic, {\it Neutrino Masses and Mixings in Gauge Models with Spontaneous Parity Violation},
  Phys.\ Rev.\ D {\bf 23}, 165 (1981).

\bibitem{seesaw3} T. Yanagida, {\it Horizontal gauge symmetry and masses of neutrinos}, Conf.  Proc.  C {\bf 7902131}, 95 (1979).

\bibitem{seesaw4} M. Gell-Mann, P. Ramond and R. Slansky, {\it Complex Spinors and Unified Theories}, Conf. Proc. C {\bf 790927}, 315 (1979) [arXiv/hep-th:1306.4669].

\bibitem{seesaw5} S. L. Glashow, {\it The Future of Elementary Particle Physics}, NATO Sci. Ser. B {\bf 61}, 687 (1980).
%

\bibitem{ChengLi} T. P. Cheng and L. F. Li, {\it Gauge theory of elementary particle physics}, Oxford university press (1994).

%
%

\bibitem{kost1} Kostelecký, V. Alan, and Matthew Mewes. "Lorentz and CPT violation in neutrinos." Physical Review D 69.1 (2004): 016005.

\bibitem{Katori} Teppei Katori, Alan Kostelecky and Rex Tayloe, {\it Global three-parameter model for neutrino oscillations using Lorentz violation},
Phys. Rev. D {\bf 74}, 105009 (2006).

\bibitem{DIAS} Diaz, Jorge S., and V. Alan Kostelecký. "Lorentz-and C P T-violating models for neutrino oscillations." Physical Review D 85.1 (2012): 016013.



\bibitem{Rosenstein} Baruch Rosenstein, Brian J. Warr and Seon H. Park, {\it Dynamical symmetry breaking in four-fermion interaction models},
Physics Reports (Review Section of Physics Letters) {\bf 205}, No. 2 (1991) 59 — 108.

\bibitem{Higashijima} Kiyoshi Higashijima, {\it Theory of Dynamical Symmetry Breaking}, Progress of Theoretical Physics Supplement No. {\bf 104}, 1991.

%
\bibitem{VictoriaPRL} M. Pérez-Victoria, {\it Exact Calculation of the Radiatively Induced Lorentz and CPT Violation in QED}, Phys. Rev. Lett. {\bf 83}, 13 (1999).


\bibitem{Miranski} V.A. Miransky, {\it Dynamical symmetry breaking in quantum field theories} (World Scientific), (1994).

\bibitem{NJL} Y. Nambu and G. Jona-Lasinio, {\it Dynamical Model of Elementary Particles Based on an Analogy with Superconductivity}, Phys. Rev. {\bf 122}, 345 (1961).

\bibitem{DLSB1} M. Gomes, T. Mariz, J. R. Nascimento, A. J. da Silva, {\it Dynamical Lorentz and CPT symmetry breaking in a 4D four-fermion model.} , Phys. Rev. D, {\bf 77}(10), p. 105002, (2008).


\bibitem{DLSB2} J. F. Assunção, T. Mariz, J. R. Nascimento, A. Y. Petrov, {\it Dynamical Lorentz symmetry breaking in a 4D massless four-fermion model.} Phys. Rev. D, {\bf 96}(6), p. 065021, (2017).

\bibitem{DLSB3} J. F. Assunção, T. Mariz, J. R. Nascimento, A. Y. Petrov, {\it Dynamical Lorentz symmetry breaking in a tensor bumblebee model.},  Phys. Rev. D, {\bf 100}(8), p. 085009, (2019).

\bibitem{DLSB4} J. F. Assunção, T. Mariz, J. R. Nascimento, A. Y. Petrov, {\it Induced Chern-Simons modified gravity at finite temperature}, J. High Energy Phys., {\bf 072} (8), p. 1-13, (2018).


\bibitem{DLSB5} B. Charneski, M. Gomes, T. Mariz, J. R. Nascimento, A. J. da Silva, {\it Dynamical Lorentz symmetry breaking in 3D and charge fractionalization}, Phys. Rev. D, {\bf 79}(6), p. 065007, (2009).


\bibitem{YuriPRD2021} Y. M. P. Gomes, {\it Dyson-Schwinger equation approach to Lorentz symmetry breaking with
finite temperature and chemical potential}, Phys. Rev. D {\bf 104}, 015022 (2021).
%

\bibitem{CFJED} Sean M. Carroll, George B. Field and Roman Jackiw, {\it Limits on a Lorentz- and parity-violating modification of electrodynamics}, Phys. Rev. D {\bf 41}, 4 (1990).

\bibitem{KanemuraPRD2014} Shinya Kanemura, Toshinori Matsui and Hiroaki Sugiyama, {\it Neutrino mass and dark matter from gauged $U(1)_{B-L}$ breaking}, Phys. Rev. D {\bf 90}, 013001 (2014).

\bibitem{ColomaJHEP2021} Pilar Coloma, M. C. Gonzalez-Garciab and Michele Maltoni, {\it Neutrino oscillation constraints on $U(1)^{\prime}$ models:
from non-standard interactions to long-range forces}, JHEP {\it 01} (2021) 114.

\bibitem{BonillaPRD2020} Cesar Bonilla, Leon M. G. de la Vega, R. Ferro-Hernandez, Newton Nath and Eduardo Peinado, {\it Neutrino phenomenology in a left-right $D4$ symmetric model}, Phys. Rev. D {\bf 102}, 036006 (2020).


%
\bibitem{PDG2020} P. A. Zyla et al (Particle Data Group), Prog. Theor. Exp. Phys. 2020, 083C01 (2020).
%

\bibitem{Mills} Geoffrey B. Mills et al., {\it Neutrino oscillation results from LSND}, Nuclear Physics B - Proceedings Supplements,
v. {\bf 91}, n. 1-3, p. 198-202 (2001).

\bibitem{ABE} K. ABE et al., {\it Solar neutrino measurements in Super-Kamiokande-IV}, Phys. Rev. D {\bf 94}, 052010 (2016).
%

\bibitem{SME2021} V. Kostelecký and N. Russell, Data tables for Lorentz and CPT violation, Rev. Mod. Phys. {\bf 83} (2011) 11 [0801.0287v13 (2021 edition)].
%

\bibitem{minos} P. Adamson et al. (MINOS+ Collaboration) {\it Precision constraints for three-flavor neutrino oscillations from the full MINOS+ and MINOS dataset}, Phys. Rev. Lett. {\bf 125}, 131802 (2020).


\bibitem{icecube} M. G. Aartsen et al. (IceCube Collaboration), {\it Determining neutrino oscillation parameters from atmospheric muon neutrino disappearance with three years of IceCube DeepCore data}. Phys. Rev. D {\bf 91}, 072004 (2015).

\bibitem{icecubePeV} M. G. Aartsen et al. (IceCube Collaboration), {\it First Observation of PeV-Energy Neutrinos with IceCube},
Phys. Rev. Lett. {\bf 111}, 021103 (2013).

\bibitem{IceCubeNature} The IceCube Collaboration, {\it Detection of a particle shower at the Glashow resonance with IceCube}, Nature {\bf 591},
220-224 (2021).


%
%
\end{thebibliography}
\end{document}